\documentclass{aa}  

\usepackage{graphicx}
\usepackage{txfonts}
\usepackage{natbib}
\usepackage{color}
\bibpunct{(}{)}{;}{a}{}{,} 
\begin{document}

   \title{The SOPHIE search for northern extrasolar planets\thanks{Based on observations collected with the SOPHIE spectrograph on the 1.93-m telescope at Observatoire de Haute-Provence (CNRS), France by the SOPHIE Consortium.}}

   \subtitle{XII. Three giant planets suitable for astrometric mass determination with \emph{Gaia}\thanks{Tables 5 to 7 are available in electronic form at the CDS via anony- mous ftp to cdsarc.u-strasbg.fr (130.79.128.5) or via http://cdsweb.u- strasbg.fr/cgi-bin/qcat?J/A+A/TBC}}

   \author{J. Rey\inst{1} \and G. H\'ebrard\inst{2,6} \and F. Bouchy\inst{1,3} \and V. Bourrier\inst{1} \and I. Boisse\inst{3} \and N. C. Santos\inst{4,5} \and L. Arnold\inst{6} \and N. Astudillo-Defru\inst{1} \and X. Bonfils\inst{7,8} \and S. Borgniet\inst{7,8} \and B. Courcol\inst{3} \and M. Deleuil\inst{3} \and X. Delfosse\inst{7,8} \and O. Demangeon\inst{3} \and R. F. D\'iaz\inst{1,11,12} \and D. Ehrenreich\inst{1} \and T. Forveille\inst{7,8} \and M. Marmier\inst{1} \and C. Moutou\inst{3,9} \and F. Pepe\inst{1} \and A. Santerne\inst{3} \and J. Sahlmann\inst{10} \and D. S\'egransan\inst{1} \and S. Udry\inst{1} \and P. A. Wilson\inst{2}}

   \institute{Observatoire Astronomique de l'Universit\'e de Gen\`eve, 51 Chemin des Maillettes, 1290 Versoix, Switzerland \and Institut d'Astrophysique de Paris, UMR7095 CNRS, Universit\'e Pierre \& Marie Curie, 98bis boulevard Arago, 75014 Paris, France \and Aix Marseille Universit\'e, CNRS, Laboratoire d'Astrophysique de Marseille UMR 7326, 13388 Marseille cedex 13, France \and 
Instituto de Astrof\'isica e Ci\^encias do Espa\c{c}o, Universidade do Porto, CAUP, Rua das Estrelas, 4150-762 Porto, Portugal \and  Departamento de F\'isica e Astronomia, Faculdade de Ci\^encias, Universidade do Porto, Rua do Campo Alegre, 4169-007 Porto, Portugal \and Observatoire de Haute Provence, CNRS, Aix Marseille Universit\'e, Institut Pyth\'eas UMS 3470, 04870 Saint-Michel-
l’Observatoire, France \and Univ. Grenoble Alpes, IPAG, F-38000 Grenoble, France \and CNRS, IPAG, F-38000 Grenoble, France \and Canada-France-Hawaii Telescope Corporation, 65-1238 Mamalahoa Hwy, Kamuela, HI 96743, USA \and European Space Agency, Space Telescope Science Institute, 3700 San Martin Drive, Baltimore, MD 21218, USA \and Universidad de Buenos Aires, Facultad de Ciencias Exactas y Naturales. Buenos Aires, Argentina \and CONICET - Universidad de Buenos Aires. Instituto de Astronom\'ia y F\'isica del Espacio (IAFE). Buenos Aires, Argentina}

   \date{Received; accepted}

 
  \abstract
{We present new radial velocity measurements for three low-metallicity solar-like stars observed with the SOPHIE spectrograph and its predecessor ELODIE, both installed at the 193 cm telescope of the Haute-Provence Observatory, allowing the detection and characterization of three new giant extrasolar planets in intermediate periods of 1.7 to 3.7 years. All three stars, HD17674, HD42012 and HD29021 present single giant planetary companions with minimum masses between 0.9 and 2.5 $M_{Jup}$. The range of periods and masses of these companions, along with the distance of their host stars, make them good targets to look for astrometric signals over the lifetime of the new astrometry satellite \emph{Gaia}. We discuss the preliminary astrometric solutions obtained from the first \emph{Gaia} data release.}


   \maketitle
%
\section{Introduction}
Measuring radial velocities (RVs) was one of the first techniques used in the search for extrasolar planets and the one that led to the discovery of the first exoplanet around a solar-like star, 51 Peg b, by \cite{Mayor1995}. More than twenty years later, the time range of radial velocity surveys is allowing us to probe the intermediate and external regions around other stars with a long-term accuracy which is high enough to detect giant planets. This is a key factor in finding planetary systems whose architecture resembles our own solar system and in understanding their formation and evolution. Also, giant planets can play an important role in the dynamics of the systems, especially affecting the inner planets. 

In October 2006, the SOPHIE consortium started a large programme to detect and characterize exoplanets \citep{Bouchy2009a}, allowing the discovery of several of them \citep[e.g.][]{Courcol2015,Diaz2016,Hebrard2016}. Some of the SOPHIE targets were first observed within the ELODIE historical programme initiated by M. Mayor and D. Queloz in 1994, this extends our baseline to over 22 years and allows a long-term characterization of the orbits. We have also benefited from the extent of this RV survey to study stellar magnetic cycles. These activity cycles can produce variations in RVs and, if not properly identified, they could lead to a false detection by mimicking a planetary signature.

For non-transiting systems, RVs only provide a minimum value for the planetary mass, since the inclination of the orbit remains undetermined\footnote{In some multi-planetary systems, there is dynamical interaction between the planets. In these cases the inclination, and therefore the true masses, can be determined by RVs alone \citep[e.g.][]{Correia2010}.}. By combining RVs and astrometry we are able to resolve this ambiguity, to fully determine the mass of the companion, and to find the parameters of the 3D orbit \citep[e.g.][]{Sahlmann2016}.\\
Unlike RVs, the amplitude of the astrometric signal will be larger for longer periods. Even though a high degree of precision is needed to detect planets, some discoveries have been claimed \citep{Muterspaugh2010} and in other cases astrometry has helped to determine the true nature of a substellar object \citep{Benedict2010,Sahlmann2011}. Many of these projects were possible thanks to the Hipparcos astrometry mission \citep{Hipparcos1997}. Hipparcos was developed by ESA and was launched in 1989. It was the first satellite mission dedicated to astrometry, and it reached a milliarcsecond accuracy, but its precision is still far from ideal for exoplanet detections. This scenario is expected to change thanks to the microarcsecond precision of the \emph{Gaia} satellite \citep{GaiaMission2016}, which was launched in December 2013. With some assumptions on planet occurrences, \cite{Perryman2014} have estimated that more than 20,000 long-period massive planets closer than 500 pc should be discovered for the nominal five-year mission of \emph{Gaia}. 
In September 2016, the first \emph{Gaia} data release (DR1) became available \citep{GaiaDR12016,Lindegren2016}. The preliminary astrometric solutions of DR1 allow us to look for hints of substellar companions and discuss the better characterization that we will achieve thanks to the synergy between RVs and astrometry after the nominal five-year mission.

We report the detection of three Jupiter-mass planets around the Sun-like stars HD17674, HD42012, and HD29021 based on ELODIE and SOPHIE RV measurements. The observations are presented in Section 2, and we characterize the host stars in Section 3. In Section 4, we describe the methodology used, as well as the Hipparcos astrometric analysis. In Section 5, we present our results and constrain the planetary parameters. We discuss \emph{Gaia} DR1 and the expected astrometric signals due to these planets in Section 6 and, finally, we present our conclusions in Section 7.

\section{Spectroscopic observations}
Two of our stars (HD17674, HD42012) were first observed with the cross-dispersed \textit{echelle} spectrograph ELODIE, mounted on the 193 cm telescope at the Haute-Provence observatory from late 1993 to mid 2006 \citep{Baranne1996}. All three targets were later observed with the fibre-fed cross-dispersed \textit{echelle} spectrograph SOPHIE, installed at the same telescope in 2006 \citep{Bouchy2009a,Perruchot2008}. SOPHIE is environmentally stabilized to provide high-precision radial velocity measurements. All SOPHIE spectroscopic observations were done using the fast reading mode of the detector and high-resolution (HR) mode of the spectrograph, providing a resolution power of R = 75000.\\
HD29021 and HD42012 were observed as part of the volume-limited SOPHIE survey for giant planets \citep[subprogram 2, or SP2; ][]{Bouchy2009a,Hebrard2016}. These targets were observed in \textit{objAB} mode, where fibre \textit{A} is used to collect the light from the star while fibre \textit{B} monitors the sky brightness variations, especially due to moonlight. A correction is applied to observations where the sky fibre shows significant moonlight pollution, following the procedure described by e.g. \cite{Bonomo2010}. Spectra are acquired with a constant signal-to-noise ratio (S/N) per pixel of around 50 at 550 nm to minimize the effects of the charge transfer inefficiency (CTI), characterized by \cite{Bouchy2009cti}. Nevertheless, some spectra with S/N lower than 25 were present in our data set, and were therefore removed. For the data we keep, we correct any remaining effects of the CTI using the empirical function described in \cite{Santerne2012}.\\
Using all 39 SOPHIE spectral orders, corrected spectra are cross-correlated with a numerical mask corresponding to the spectral type of the observed stars in order to obtain the cross-correlation functions (CCFs). We apply a Gaussian fit on the CCFs to obtain the radial velocities \citep{Baranne1996,Pepe2002}, and also to extract the CCF parameters (FWHM and contrast) and the bisector span (BIS), as described by \cite{Queloz2001}. For the RV uncertainties, we use $1\sigma$ values.\\
To account for the instrumental drift of the SOPHIE spectrograph, wavelength calibrations are made every 2-3 hours during the night and this value is interpolated for each exposure in \textit{objAB} mode only when the time between calibrations is not greater than four hours, resulting in a correction of the order of 1 $ms^{-1}$.\\
HD17674 was part of the follow-up of ELODIE long periods \citep[subprogram 5, or SP5;][]{SOPHIEV,Bouchy2009a,Bouchy2016}. Subprogram 5 needs a higher level of precision to detect giant long-period planets (which are expected to have lower RV semi-amplitudes than SP2 planets) and to constrain the offset between ELODIE and SOPHIE. All measurements for this target were acquired using the \textit{thosimult} mode, where the stellar spectrum from fibre \textit{A} is recorded simultaneously with a thorium-argon calibration from fibre \textit{B} to estimate the instrumental drift of the spectrograph. Since the information regarding moonlight pollution is not available in this mode, we checked the values of the barycentric Earth radial velocity and the closeness of the Moon to ensure our spectra were not polluted.\\
To achieve the expected precision, our analysis did not include ELODIE spectra with S/N lower than 50 or SOPHIE spectra with S/N lower than 80. We did not include SOPHIE spectra with unusual ThAr flux since they could be polluted and could affect our RV determination. SOPHIE spectra in \textit{thosimult} mode were reduced using the pipeline described by \cite{Bouchy2009a}. The cross-correlation process is the same as described above.

In June 2011 (BJD = 2455730), the SOPHIE spectrograph was upgraded. In order to minimize a systematic effect produced by a seeing change at the fibre input, referred to as \emph{seeing effect} \citep{Boisse2010,Boisse2011a,Diaz2012}, a piece of octagonal-section fiber was installed in the SOPHIE fiber link \citep{Bouchy2013}. Therefore, we distinguish two different datasets: SOPHIE and SOPHIE+, before and after BJD = 2455730, respectively. We applied a partial correction to the SOPHIE data set to account for the seeing effect, which is done by measuring the difference between the RVs on the blue and red parts of the spectrum and using this value to decorrelate the velocities measured using the entire detector \citep{Bouchy2013, Diaz2012}. Due to the scrambling properties before BJD = 2455730, we quadratically added a systematic RV uncertainty of 5 $ms^{-1}$ on the SOPHIE data set. This value corresponds to the RV precision measured on stable stars at that time \citep{Bouchy2013}. \\
Finally, the SOPHIE spectrograph is affected by additional instrumental drifts (e.g. due to interventions in the instrument, lamp replacements). With the precision of SOPHIE+ this effect became detectable, and is well characterized thanks to the systematic observation of RV constant stars. The long-term variations of the zero point of the instrument identified by \cite{Courcol2015} added up to about $\pm10$ $ms^{-1}$ over 3.5 years. To correct for these effects, they used data from the monitored constant stars HD185144, HD9407, HD221354, and HD89269A, and from 51 other targets from their sample with at least ten measurements. Then they recursively built a RV constant master with these two data sets, starting with the constants and recursively adding corrected targets (from the 51-star sample) with a root mean square lower than a certain threshold (3 $ms^{-1}$). The complete process is detailed in \cite{Courcol2015}. We applied the RV constant master correction to all of our SOPHIE+ data, achieving a RV precision in HR mode close to 2 $ms^{-1}$.

\section{Spectral analysis}

\subsection{Stellar parameters} A spectroscopic analysis was performed on the combined high-resolution spectra obtained with SOPHIE. Exposures polluted by light from the Moon are not included on the combined spectrum. For HD29021, 60 spectra were used, while for HD42012, 20 were used. The average S/N of the spectra used for both targets was 50. For HD17674, two spectra without simultaneous thorium-argon calibration were taken from the SOPHIE archive for this analysis. These two observations were made on October 25 and 26, 2009, in \emph{objA} mode and with exposure times of 300 seconds, reaching a S/N of around 100.\\
Effective temperature $T_{eff}$, surface gravity $\log g$ and metallicity [Fe/H] were accurately derived using the method described in \cite{Santos2004} and \cite{Sousa2008}. The procedure is based on the equivalent widths of the FeI and FeII lines and the iron excitation and ionization equilibrium, which is assumed to be in local thermodynamic equilibrium. Stellar masses $M_*$ are derived using the spectroscopic parameters as input for the calibration of \cite{Torres2010} with a correction following \cite{Santos2013}. Errors were estimated from 10 000 random values of the stellar parameters within their error bars and assuming a Gaussian distribution. The stellar ages were derived by interpolation of the PARSEC\footnote{http://stev.oapd.inaf.it/param} \citep{Bressan2012} tracks using the method described by \cite{daSilva2006}. Estimated errors only include the uncertainties intrinsic to this analysis, so for all computations requiring stellar masses (e.g. the masses of the companions), we decided to use a conservative $10\%$ error. Finally, we estimated the projected stellar rotational velocities, $v \sin i$, from the SOPHIE CCF as described by \cite{Boisse2010}.
All the obtained values are listed in Table~\ref{StarPar}.

\begin{table*}[ht]
\caption{Target characteristics and summary of observations.}              
\label{StarChar}      
\centering                                      
	\begin{tabular}{l c c c c c c c c c}          
	\hline\hline                        
	Target  &	RA	&	DEC	&	V 	&	B-V	&	Spectral		&	$\pi$ & Distance	&	Time Span		&	Nmeas	\\   
	Name  &	(J2000)	&	(J2000)	&		& 			&	Type		& [mas]	&	[pc]	&	[years]	&	Elodie/Sophie	\\
	\hline  
	HD17674	& 02:51:04.3 & +30:17:12.3 & 7.56 & 0.58 & G0V  & $22.46\pm0.23^{(1)}$ & $44.5\pm0.8$& 18.37 & 8/93 \\
	HD42012  & 06:09:56.5 & +34:08:07.1 & 8.44 & 0.79  & K0  & $26.94\pm0.90^{(2)}$ & $37.1\pm1.2$ & 8.24 & 1/31    \\                        
	HD29021  & 04:37:52.2 & +60:40:34.3 & 7.76 & 0.71 & G5  & $32.64\pm0.30^{(1)}$ & $30.6\pm0.4$ & 4.41 & 0/66 \\
	\hline     
	\end{tabular}	

\begin{flushleft}
		\begin{small}
$^{(1)}$ Parallaxes from \emph{Gaia} DR1. Nominal uncertainties in the parallax, but adding a 0.3 mas systematic error is recommended \citep[see][]{Lindegren2016}.\\
$^{(2)}$ Parallax from Hipparcos archive, since no value is available from \emph{Gaia}.
		\end{small}
\end{flushleft}
\end{table*}

\begin{table*}[ht]
\caption{Stellar parameters.}              
\label{StarPar}      
\centering                                      
	\begin{tabular}{l c c c c c c c c}          
	\hline\hline                        
	Target & $T_{eff}$	& log g & Fe/H & M$_*$			& R$_*$ 		  & Age & $v \sin i$ &  log $R'_{HK}$ \\
	Name & [K]			& [cgs] & [dex] &  [M$_\odot$] & [R$_\odot$]  & [Gyr]&[km/s]	 & \\
	\hline                
	HD17674 & $5904\pm22$ & $4.34\pm0.03$ & $-0.16\pm0.02$ & $0.98\pm0.10^{(1)}$ & $1.18\pm0.1$ & $8.4\pm0.6$ &$2.3\pm1.0$ & $-4.91\pm0.17$\\
	HD42012 & $5405\pm45$ & $4.45\pm0.07$ & $-0.09\pm0.08$ & $0.83\pm0.08^{(1)}$ & $0.82\pm0.08$ & $4.1\pm3.6$ & $2.2\pm1.0$& $-5.00\pm0.13$ \\
	HD29021 & $5560\pm45$ & $4.44\pm0.03$ & $-0.24\pm0.02$ & $0.85\pm0.08^{(1)}$ & $0.85\pm0.09$ & $7.4\pm3.1$ & $2.7\pm1.0$& $-5.00\pm0.14$ \\       
	\end{tabular}	
	
\begin{flushleft}
		\begin{small}
$^{(1)}$ A conservative $10\%$ error is used.\\
		\end{small}
\end{flushleft}
\end{table*}

\subsection{Bisector span analysis}   
In our analysis we studied the behaviour of the bisector span. Blended stellar systems could mimick the RV signature of a planet around a star, but a planet will only produce a shift of the stellar lines. Blended systems, on the other hand, will change the shape of the lines and this could be reflected in the bisector velocity span that we obtain from the CCFs. Bisector variations can also be produced by stellar activity over the timescale of the rotational period of the star \citep[see e.g.][]{Boisse2011a} and also by stellar activity cycles.
   We examined the bisector for all of our candidates to look for correlations with the RV measurements or their residuals that could reveal the presence of a companion polluting the peak of the primary star or signs of stellar activity. None of our targets present correlations between the bisector span and the RVs.
   Regarding the behaviour of the bisector span during time, no significant variations were detected and all values lie within $3\sigma$.

\subsection{Activity indicators}
The stellar activity level was estimated on each spectrum by measuring the emission in the core of the Ca II H and K lines using the calibration described in \cite{Boisse2010}. The mean $\log R'_{HK}$ values and standard deviation obtained from SOPHIE spectra are given in Table~\ref{StarPar}.
We decided to extend our analysis by adding a second known stellar activity indicator, H$_\alpha$, for two main reasons: first, for ELODIE spectra the Ca II H and K lines fall in the first orders of the wavelength range where the S/N is very poor. Second, due to to the lower S/N of the SP2 SOPHIE spectra, we also get a typically low flux in the Ca II H and K region. Nevertheless, for SP2 the average $\log R'_{HK}$ value is a good enough indicator of the mean stellar activity level.\\
At the spectral location of H$_\alpha$, both ELODIE and SOPHIE have a higher instrumental response, which allows us to better trace the long-term stellar activity induced by the increase and decrease of active regions. The indicator H$_\alpha$ has proven to be a good chromospheric indicator even though it forms at lower altitudes in the stellar atmosphere than the Ca II H and K lines. In addition, \cite{Gomesdasilva2014} showed that around 23\% of the FGK stars in their sample showed strong correlations (positive or negative) between H$\alpha$ and calcium indices. Because of this, H$\alpha$ has been used by several authors \citep[e.g.][]{Pasquini1991,Montes1995,Cincunegui2007a,Cincunegui2007b,Robertson2013, Robertson2014,Neveu2016}. The parameters used for our H$_\alpha$ analysis can be found in \cite{Diaz2016}.\\
It is worth mentioning that SP5 aims to detect Saturn and Jupiter analogues; therefore, it is important to disentangle long-period planets from long-term stellar activity due to magnetic cycles. Having reliable values for both the $S_{index}$ and H$_\alpha$ indices is another reason to ask for higher S/N for this programme. A longer baseline, like the one we have for H$_\alpha$, will allow us to better identify long-term activity.
\\
The periodograms of both $S_{index}$ and H$_\alpha$ were compared with the estimated stellar rotational periods \citep{Noyes1984} to detect the short timescale activity component, but no significant peaks were found near these values ($P_{rot}\sim$15, 40, and 32 days for HD17674, HD42012, and HD29021, respectively). \\

\begin{figure}
   \centering
   \includegraphics[width=9cm]{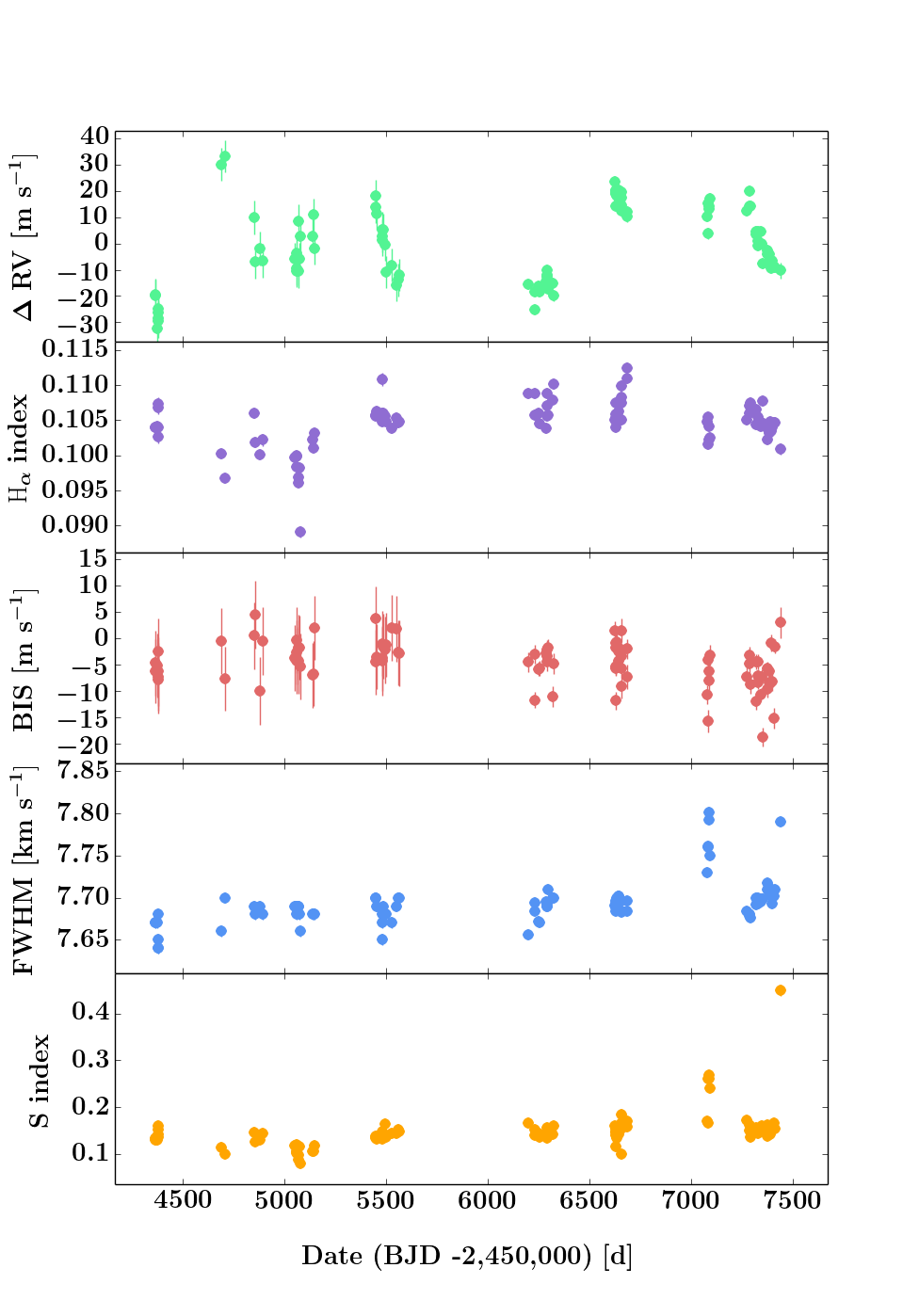}
      \caption{SOPHIE activity indicators for HD17674.}
         \label{actHD17}
   \end{figure}

\begin{figure}
   \centering
   \includegraphics[width=9cm]{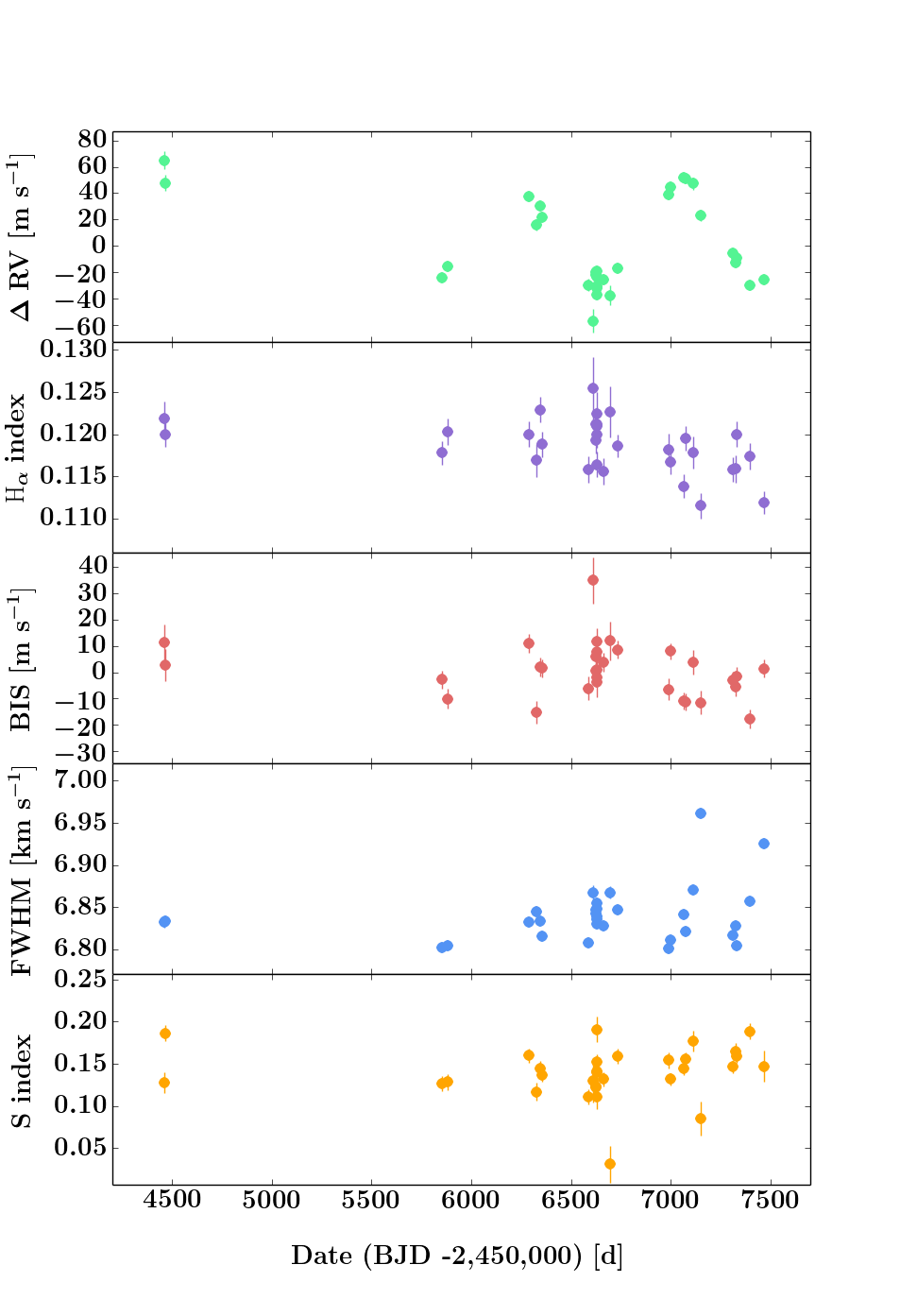}
      \caption{SOPHIE activity indicators for HD42012.}
         \label{actHD42}
   \end{figure}

\begin{figure}
   \centering
   \includegraphics[width=9cm]{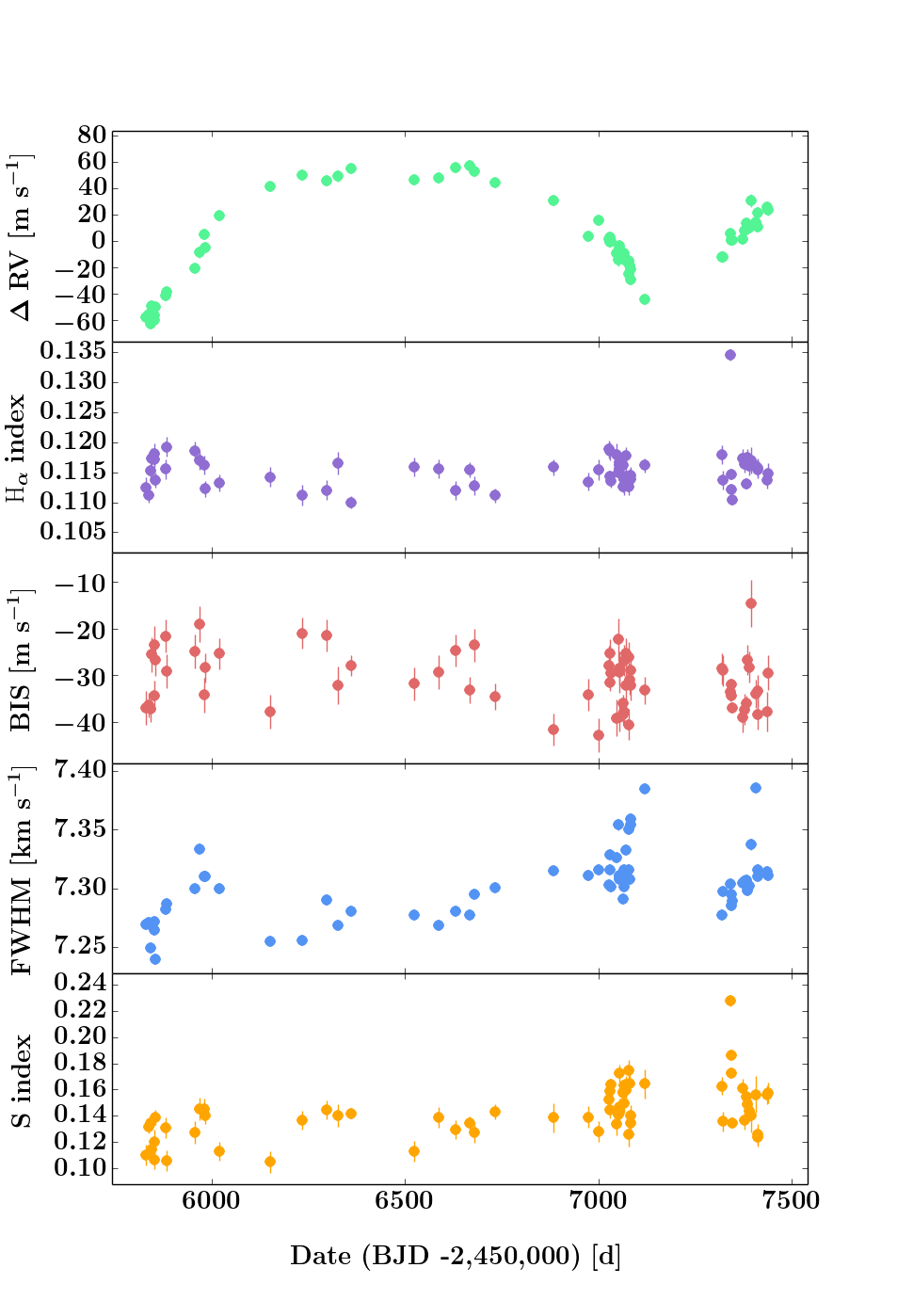}
      \caption{SOPHIE activity indicators for HD29021.}
         \label{actHD29}
   \end{figure}

In Figs. \ref{actHD17} - \ref{actHD29}, we show the evolution of the H$_\alpha$ index along with other activity indicators for our three targets. For HD17674 (Fig. \ref{actHD17}), we see a significant long-term variation in H$_\alpha$ of around 3 000 days, but this variation is not seen in any of the other indicators. We find a moderate correlation between the S index and the FWHM (Pearson correlation coefficient of 0.7). For HD42012 (Fig. \ref{actHD42}), no significant variations were detected and a moderate correlation (Pearson correlation coefficient of 0.6) was detected between the H$_{\alpha}$ index and the BIS. No significant variations or correlations between the indices were found for HD29021 (Fig. \ref{actHD29}).\\
Since stellar activity is not affecting the RVs or the residuals for our targets, no corrections were necessary for our data. 

\section{Data analysis}
\subsection{Method}
ELODIE and SOPHIE radial velocities were fitted using the tools available in the Data and Analysis Center for Exoplanets (DACE\footnote{ The DACE platform is available at http://dace.unige.ch.}) developed by the National Centre of Competence in Research \emph{PlanetS}. A preliminary solution is found using a periodogram of RVs which, for all three targets, shows clear peaks at the periods of each orbit. The analytical method used for computing the orbital parameters from the periodogram of RVs is described in detail in \cite{Delisle2016}. The results of this first analytical approximation are used as uniform priors for a Markov chain Monte Carlo (MCMC) model that we used to determine our final parameters and errors. The algorithm used for the MCMC is described in detail in \cite{Diaz2014,Diaz2016harps}. For the MCMC, we fit five parameters of the Keplerian orbit and a linear drift (to set the intervals in Table \ref{PlanPar}). For HD17674, we also fit the instrumental offset between SOPHIE and ELODIE. For HD42012 the offset is fixed using \cite{SOPHIEV} since we only use the ELODIE point to constrain any possible long-term drift. We did not model the stellar jitter because these targets are non-active G and K dwarfs and because no correlations with the activity indicators were found. For the orbital parameters, listed in Table \ref{PlanPar}, we used the mode estimate and the errors corresponding to the 68.3\% confidence intervals. For the detection limits mentioned in sections 5.1 through 5.3, in the residuals we injected planets in circular orbits at different trial periods. Using a generalized Lomb-Scargle periodogram, an injected planet is considered detected when the false alarm probability is lower than $1\%$.

\subsection{Astrometric analysis}
We analysed the astrometric data available from the Hipparcos mission \citep{Hipparcos1997} to search for signatures of orbital motion. We employed the new Hipparcos reduction \citep{vanLeeuwen2007} and followed the procedure described in \cite{Sahlmann2011b}. 
No significant orbital signals were detected in the Hipparcos astrometry for these stars. As was done in previous works \citep[e.g.][]{Diaz2012} by using the Hipparcos data and the parameters found from the RV analysis, we were able to set upper mass limits for two of the companions. For HD17674, the upper mass limit of the companion is 89 $M_{Jup}$, and for HD42012 it is 67 $M_{Jup}$. No constraints could be established for HD29021 since the orbital period is not covered by Hipparcos data.

\section{Results}

\subsection{HD17674}
\begin{figure}[h]
   \centering
   \includegraphics[width=8cm]{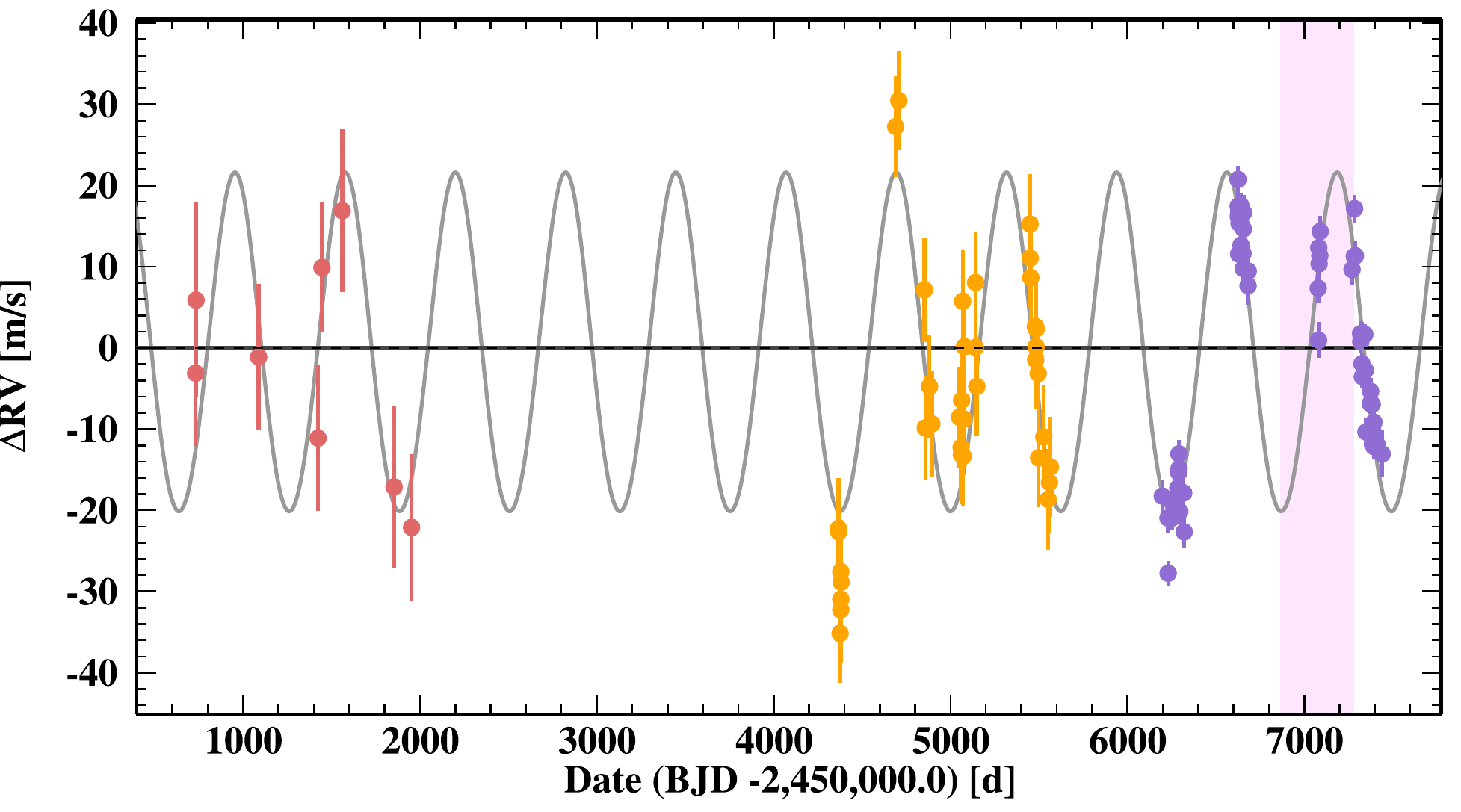}
   \includegraphics[width=8cm]{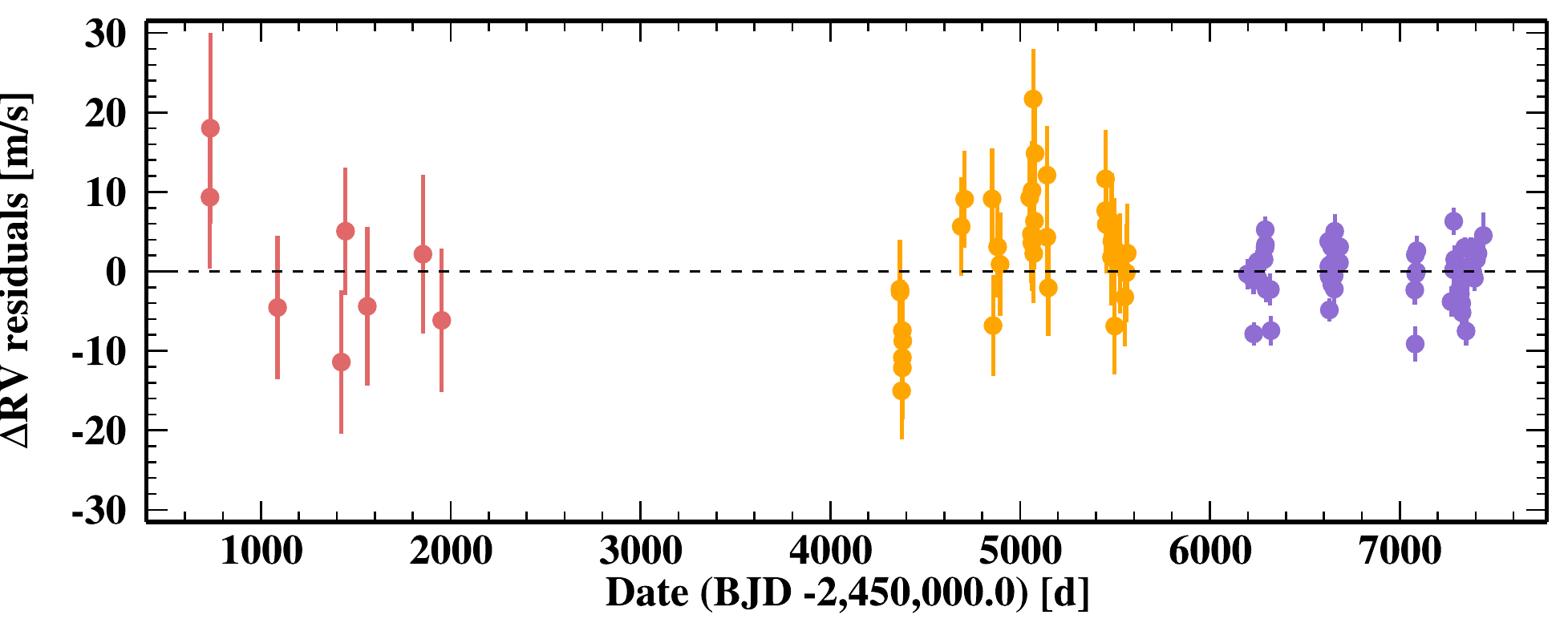}
   \includegraphics[width=8cm]{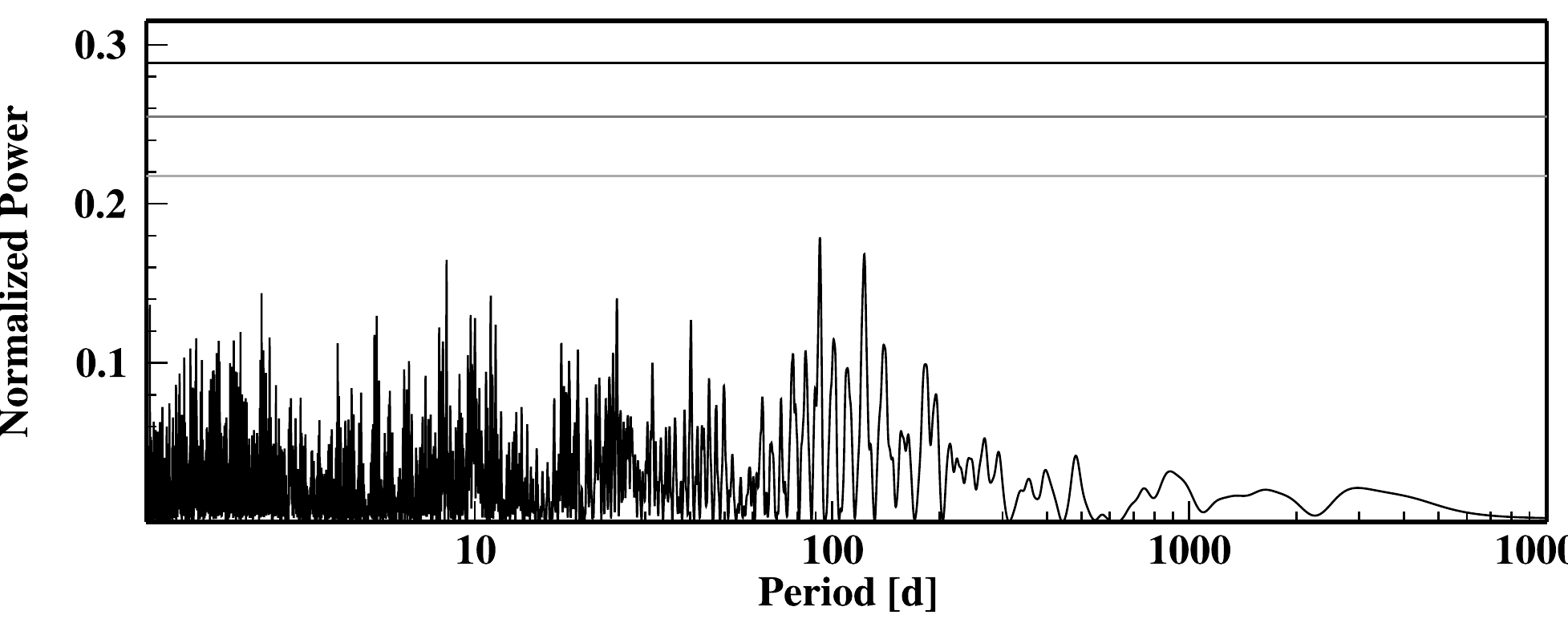}
      \caption{Radial velocity curve (top) and residuals (middle) of HD17674. ELODIE data shown in red, SOPHIE in orange, and SOPHIE+ in purple. The region in pink indicates the \emph{Gaia} observation dates for DR1. Generalized Lomb-Scargle (GLS) periodogram (bottom) of the radial velocities after subtraction of the planetary orbit. False alarm probability lines are plotted for 50\%, 10\%, and 1\%.}
         \label{FigHD17RV}
   \end{figure}

HD17674 was observed with both ELODIE and SOPHIE spectrographs with, respectively, 8 and 93 measurements for a total time span of over 18 years. One ELODIE point was excluded from the analysis, due to low S/N. For the same reason, nine SOPHIE points were not considered. Due to abnormal flux levels of the thorium-argon lamp, eight measurements were excluded. The 18 points that were excluded do not significantly change the solution we find, and would only increase the value of $\sigma_{(O-C)}$.\\
HD17674 is a V = 7.56 mag G0V star with a mass of $M_*$ of $0.98 \pm 0.10 M_{\odot}$ located at 44.5 parsecs from the Sun.  The parameters of the Keplerian fit are listed in Table \ref{PlanPar} and the orbital solution is shown in Fig. \ref{FigHD17RV}. \\
Our analysis indicates the presence of a companion of minimum mass M$_c \sin i=0.9$ M$_{Jup}$ on a 1.7-year orbit (see Fig. \ref{FigHD17RV}). We find a non-significant eccentricity with an upper $3\sigma$ limit of $e=0.13$. No significant long-term drift was detected in the data, and we can exclude giant planets more massive than M$_c \sin i=2.4$ M$_{Jup}$ on circular orbits shorter than 37 years. In Table \ref{PlanPar} we included the 99\% confidence intervals for this value. This target was also observed in high cadence to discard the presence of an inner companion. The periodogram of the residuals can be seen in Fig. \ref{FigHD17RV} and does not show any significant signals. We can exclude the presence of giant planets more massive than 0.05 and 0.1 M$_{Jup}$ with periods shorter than 10 and 100 days, respectively. The dispersion of the residuals (Table \ref{PlanPar}) corresponds well to the expected precision for each instrument. \\
The instrumental offset between ELODIE and SOPHIE was left as a free parameter and was adjusted to $57\pm6$ $ms^{-1}$, in agreement with the expected offset at $46\pm23$ $ms^{-1}$ \citep{SOPHIEV}.\\
In addition, we followed the procedure described by \cite{Diaz2016} to calculate the luminosity of HD17674 and to check whether the planet was in the habitable zone. We used the habitable zone calculator \footnote{http://depts.washington.edu/naivpl/sites/default/files/hz.shtml} based on the work by \cite{Kopparapu2013}. We chose the Runaway Greenhouse limit for our inner limit and the Maximum Greenhouse limit as the outer one. This yields a habitable zone between 1.2 AU and 2.1 AU. With a semi-major axis of 1.42 AU and an almost circular orbit, HD17674 b falls well inside this region. This is important for habitability if exomoons can be detected with future techniques around this kind of target. Since HD17674 is an evolved star, it is possible that the detected planet was outside the habitable zone when the star was still in the main sequence.

\subsection{HD42012}
\begin{figure}[h]
   \centering
   \includegraphics[width=8cm]{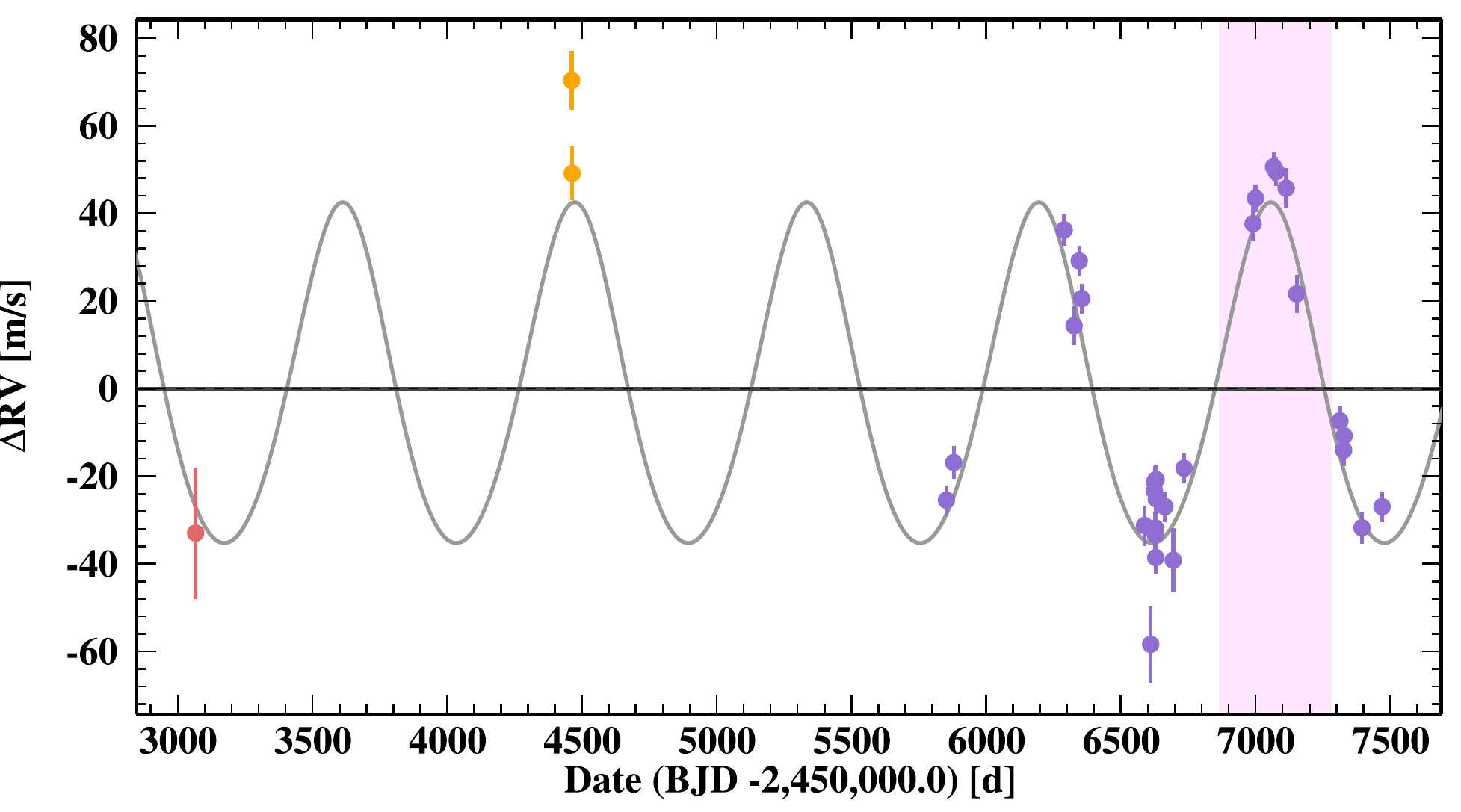}
   \includegraphics[width=8cm]{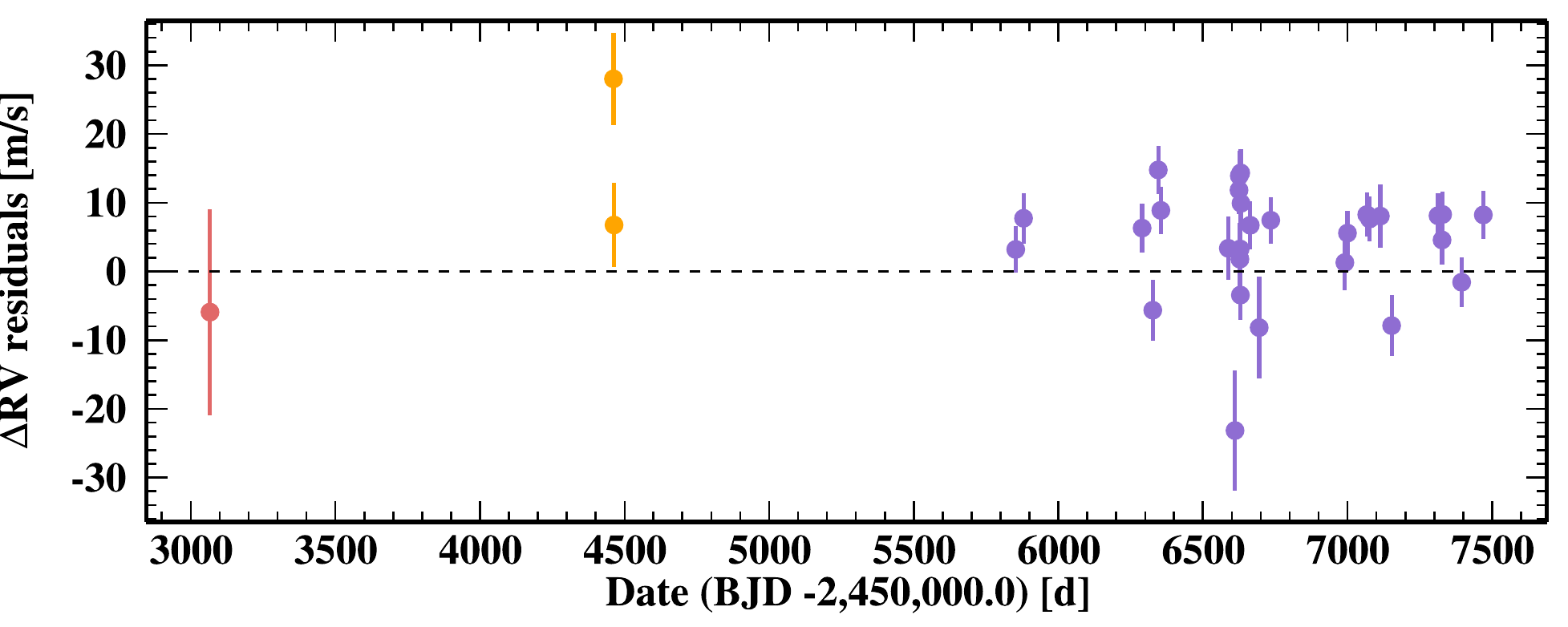}
   \includegraphics[width=8cm]{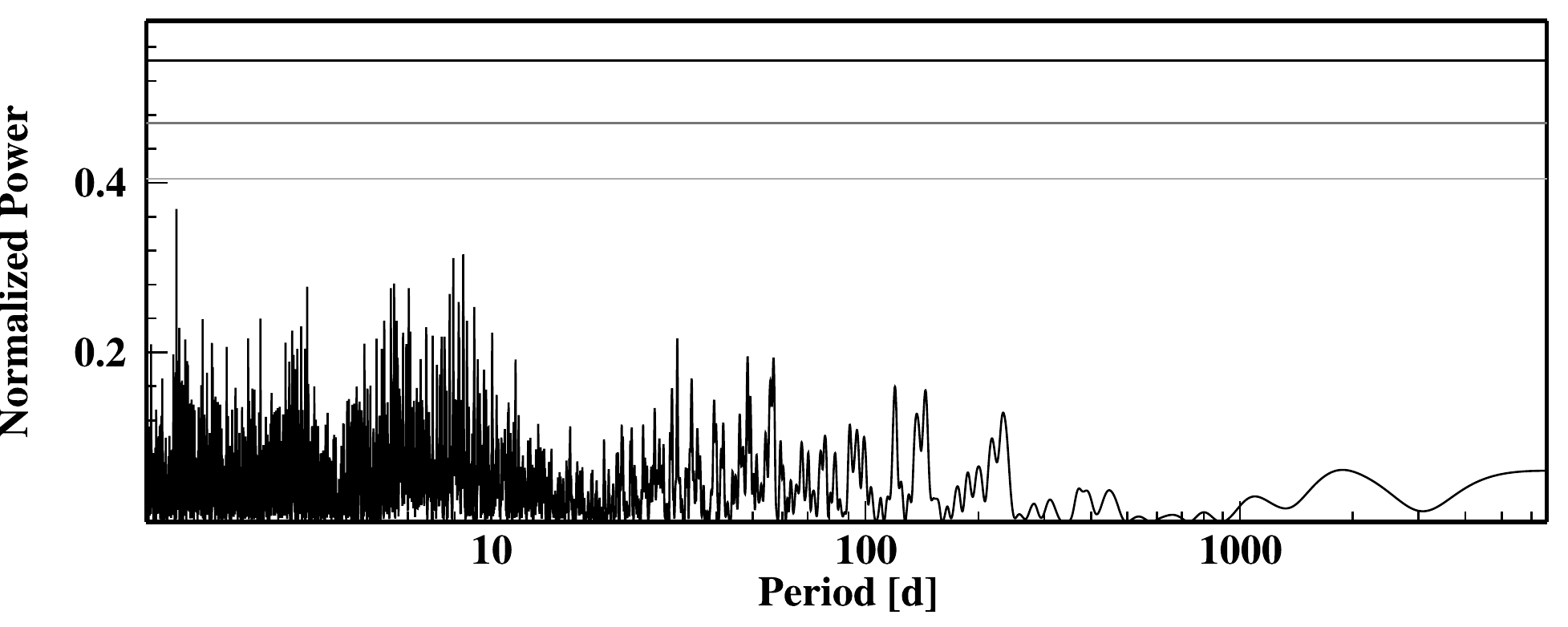}
      \caption{Radial velocity curve (top) and residuals (middle) of HD42012. GLS periodogram (bottom) of the RVs after subtraction of the planetary orbit. Colours and symbols are the same as in Fig.\ref{FigHD17RV}.}
         \label{FigHD42RV}
   \end{figure}

HD42012 was observed once with ELODIE in \textit{thosimult} mode in March 2004, twice with SOPHIE before the upgrade, and 29 times with SOPHIE+ for a total duration of 12 years. No measurements were excluded from the analysis. The star is a V = 8.44 mag K0 star with a mass of $M_*$ of $0.83 \pm 0.08 M_{\odot}$ located at 37.1 parsecs from the Sun.\\ The Keplerian fit (see Fig. \ref{FigHD42RV}) indicates the presence of a companion of minimum mass M$_c \sin i=1.6$ M$_{Jup}$ on a 2.3-year orbit. We find a non-significant eccentricity with an upper $3\sigma$ limit of $e=0.20$. All parameters are listed in Table \ref{PlanPar}.\\
The offset between ELODIE and SOPHIE was fixed using \cite{SOPHIEV}, which yields an absolute value of $134\pm23$ $ms^{-1}$. We decided not to fit this parameter as we did for HD17674; in this case, only one ELODIE point was available and thus the offset fitting would not be informative. Thanks to this additional point we can rule out any significant long-term drift. We can exclude giant planets more massive than M$_c \sin i=2.1$ M$_{Jup}$ on circular orbits shorter than 16 years. In Table \ref{PlanPar} we included the 99\% confidence intervals for this value. The periodogram of the residuals can be seen in Fig. \ref{FigHD42RV} and it shows that no other significant signals are present in the data. The dispersion of the residuals is higher than expected for the precision of the instruments, which can indicate for example that we underestimated the activity level or there is an additional undetected companion. Nevertheless, we can exclude the presence of giant planets more massive than 0.06 and 0.02 M$_{Jup}$ with periods shorter than 10 and 100 days, respectively. \\
Following the same habitable zone analysis, this planet lies beyond the Maximum Greenhouse limit of HD42012.
 
\begin{table*}[ht]
\caption{Orbital parameters}              
\label{PlanPar}      
\centering                                      
	\begin{tabular}{l c c c}          
	\hline\hline                        
	Parameters 					& HD17674 			& HD42012 				& HD29021 \\    
	\hline                                   
	    P [days]$^{(1)}$				& $623.8^{+1.6}_{-1.5} $ 			& $857.5^{+6.2}_{-6.3}$	& $1362.3^{+4.6}_{-4.0}$ 	 \\  
	    K [ms$^{-1}$]$^{(1)}$	 	&	$21.1^{+0.6}_{-0.5}$ 				& $39.0\pm0.9$			& $56.4\pm0.9$ \\     
	    e $^{(1)}$					& $<0.13$					& $<0.2$		&	$0.459 \pm 0.008$ 	 \\  
	    $T_{p}$	[RJD]$^{(1)}$			& $2455904.8\pm 3$ & $2455386.2\pm 10$ & $2455823.9^{+6.1}_{-5.7}$\\
	    $\omega$ [deg]$^{(1)}$		& - & - & $179.5\pm2.0$\\
	    $a$ [AU]$^{(1)}$				&	$1.42^{+0.04}_{-0.05}$ 			& $1.67^{+0.05}_{-0.06}$			&  $2.28^{+0.07}_{-0.08}$ 		\\  
	    M$_c \sin i$ [M$_{Jup}$]$^{(1)}$	 & $0.87^{+0.07}_{-0.06}$	& $1.6\pm0.1$			&	$2.4\pm0.2$ 		\\  
	    ELODIE $\sigma$(res) [ms$^{-1}$] & 8.24 & 7.73 & -\\
	    SOPHIE $\sigma$(res) [ms$^{-1}$] & 8.23 & 20.07 & -\\
	    SOPHIE+ $\sigma$(res) [ms$^{-1}$] & 3.12 & 8.61 & 3.93\\
	    RV drift [m s$^{-1}$ yr$^{-1}$]$^{(2)}$ & [-0.7;+2.2]  & [-3.9;-0.1]  & [-3.2;+0.6]\\
	\hline     
	\\                         
	\end{tabular}
	\begin{flushleft}
		\begin{small}
		$^{(1)}$Uncertainties correspond to the 68.3\% confidence intervals.\\
       $^{(2)}$ 99\% confidence intervals.
		\end{small}
\end{flushleft}
	\end{table*}
   
   \subsection{HD29021}
\begin{figure}[h]
   \centering
   \includegraphics[width=8cm]{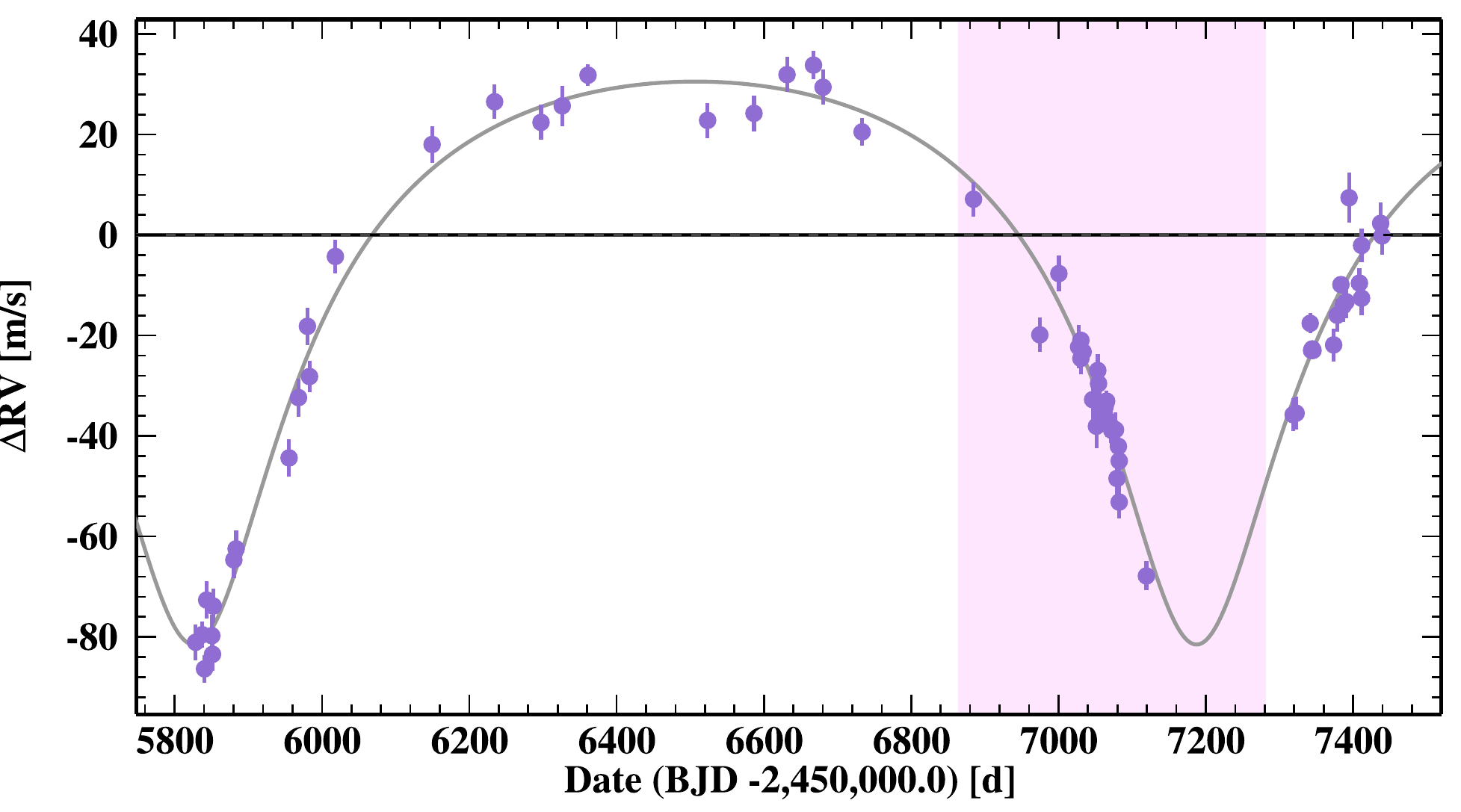}
   \includegraphics[width=8cm]{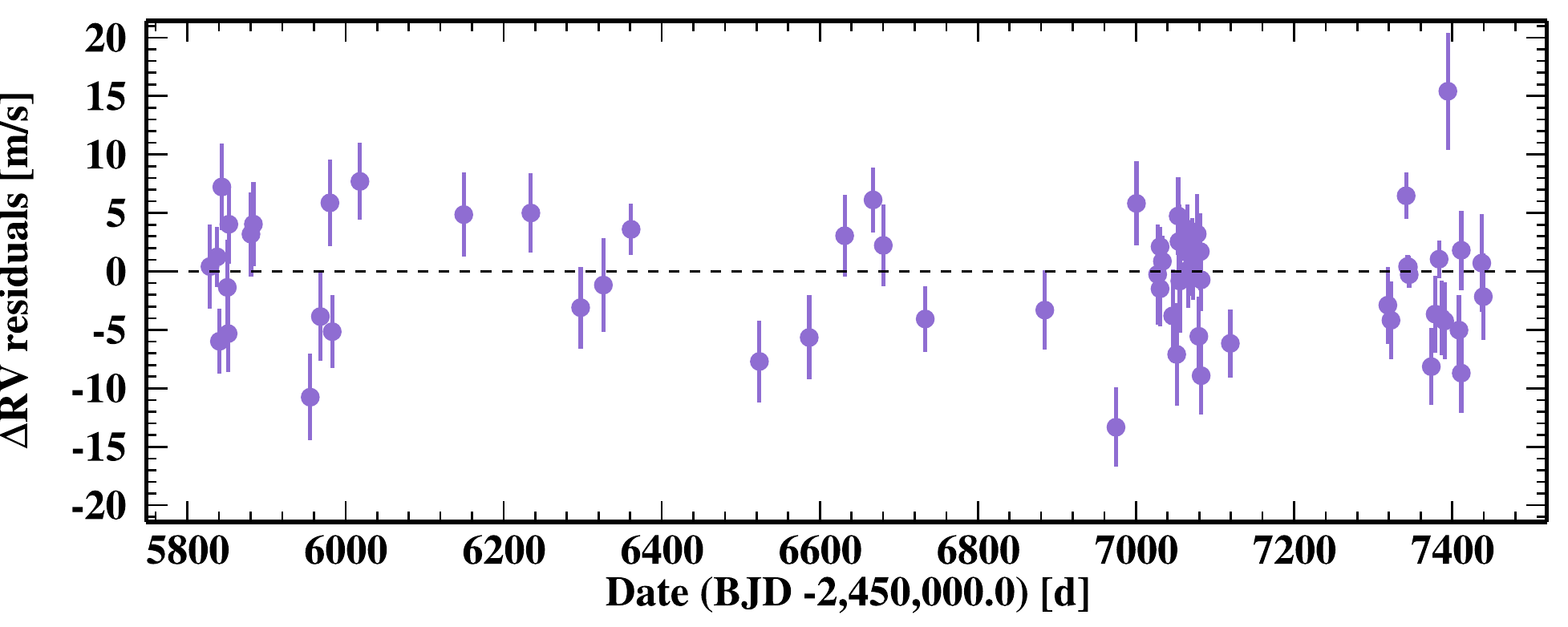}
   \includegraphics[width=8cm]{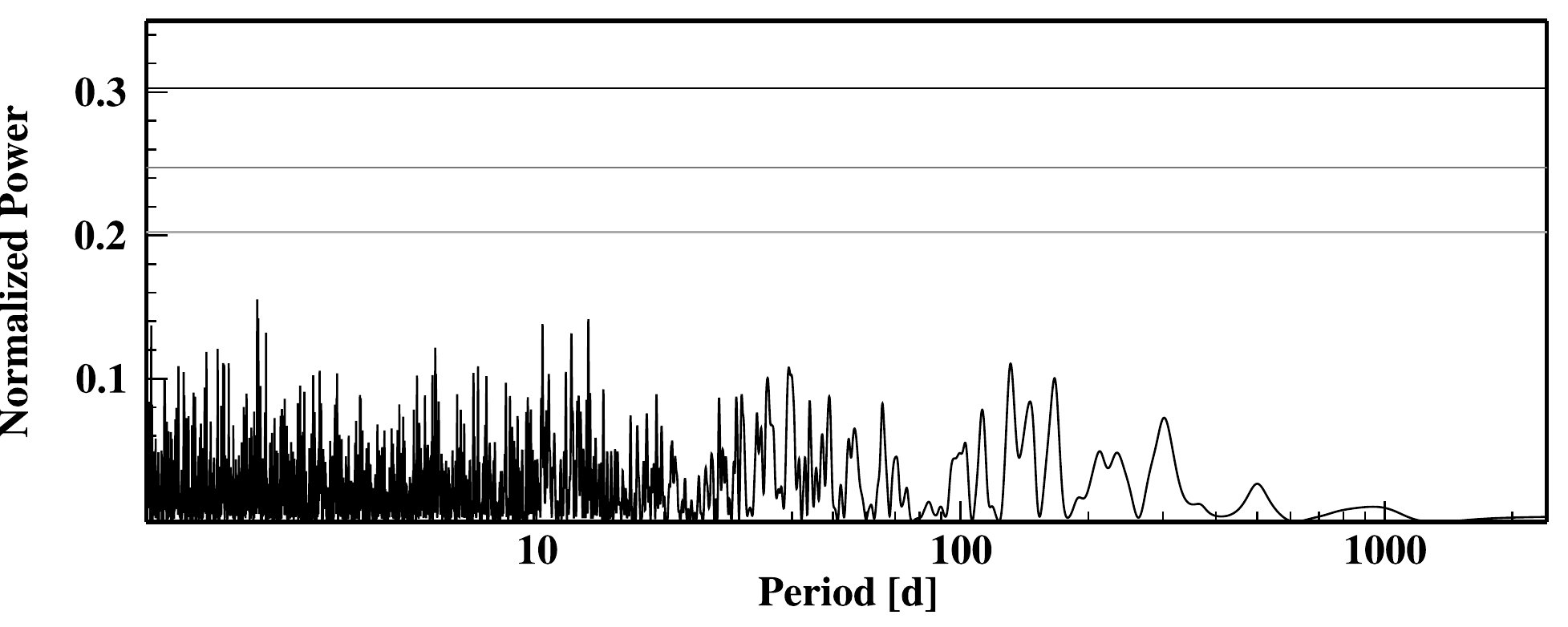}
      \caption{Radial velocity curve (top) and residuals (middle) of HD29021. GLS periodogram (bottom) of the RVs after subtraction of the planetary orbit. Colours and symbols are the same as in Fig.\ref{FigHD17RV}.}
         \label{FigHD29RV}
   \end{figure}

HD29021 was only observed with SOPHIE+ with 66 measurements spanning almost 4.5 years. Two measurements were excluded from the analysis owing to low S/N. Nevertheless, their presence did not affect the resulting orbit in a significant way. The star is a V = 7.76 mag G5 star with a mass of $M_*$ of $0.85 \pm 0.08 M_{\odot}$ located at 30.6 parsecs from the Sun. \\
The solution we find (see Fig. \ref{FigHD29RV}) corresponds to a companion of minimum mass M$_c \sin i=2.4$ M$_{Jup}$ on a 3.7-year  eccentric orbit. All orbital parameters are listed in Table \ref{PlanPar}. No significant long-term drift was detected in the data, and we can exclude giant planets more massive than M$_c \sin i=1.2$ M$_{Jup}$ on circular orbits shorter than 9 years. In Table \ref{PlanPar} we included the 99\% confidence intervals for this value.  The periodogram of the residuals is shown in Fig. \ref{FigHD29RV} and, as in the previous cases, no other significant signals are present in the data. The dispersion of the residuals is also within the expected levels for SOPHIE+. We can exclude the presence of giant planets more massive than 0.1 and 0.3 M$_{Jup}$ with periods shorter than 10 and 100 days, respectively. \\
Following the same analysis as before, this planet lies beyond the outer limit of the habitable zone of HD29021.
 
\section{Gaia astrometry}

Due to the characteristics of our planets and their host stars, we were also interested in studying the expected astrometric detection with \emph{Gaia} and its microarcsecond precision. In this section, we will discuss four different approaches.\\
In astrometry, several factors will determine the detectability of a planet, the number of field crossings of the star and their distribution, the orbital period of the planet, the orbital phase of elliptical orbits, and of course the mission duration. 
The periods for which \emph{Gaia} will be most efficient are $0.2 \lesssim P \lesssim 6$ years \citep{Perryman2014}, so our periods between 1.7 and 3.7 years are well in this interval, which means that at least one full orbit will be covered during the five-year mission of \emph{Gaia}.
\cite{Casertano2008} estimated from their numerical double-blind simulations that a planet will be detected by \emph{Gaia} with a small number of false positives, if the astrometric signal-to-noise per field crossing $S/N_{1meas}>3$ and  the period $P\leq5$ years. The value $S/N_{1meas}$ will be the first estimator that we will discuss. It is defined as $S/N_{1meas}\equiv\alpha/\sigma_{fov}$, where $\sigma_{fov}$ is the accuracy of a single field-of-view crossing and $\alpha$ is the astrometric signature given by 

\begin{equation} \label{eq:alpha}
\alpha = \left(\frac{M_p}{M_{Jup}}\right)   \left(\frac{M_\star}{M_\odot}\right)^{-1}   \left(\frac{a_p}{1 AU}\right)  \left(\frac{d}{1 pc}\right)^{-1}*954.6 \quad  \mu as,
\end{equation}
\\

for a circular orbit. Values for our three planets are listed in Table \ref{GaiaPar}. As in \cite{Perryman2014}, we take the latest estimates of $\sigma_{fov}=34.2$ $\mu as$, which is constant for targets with magnitude G$\leq$12 in the G band of \emph{Gaia}. 
 Using the detection criterion of \cite{Casertano2008} and $\sigma_{fov}=34.2$ $\mu as$, a planet around a star with magnitude G$\leq$12 and period $P\leq5$ years would be detectable by \emph{Gaia} if $\alpha \gtrsim 103$ $\mu as$. According to this, HD29021 b could be detected by \emph{Gaia} alone. Since this target is eccentric, in Table \ref{GaiaPar} we gave an estimate of the maximum value of $\alpha$ when the planet is at apastron. In Fig. \ref{FigHD29RV} we can also see which part of the orbit was covered by \emph{Gaia} DR1. For their final number of detections, \cite{Perryman2014} consider that a more liberal threshold of $S/N_{1meas}>2$ is a reasonable approximation. In this case, we would need $\alpha \gtrsim 68$ $\mu as$ and hence HD42012 b could also be detected.\\

As \cite{Perryman2014} indicate in their work, while $S/N_{1meas}$ provides some indication of planet detection numbers, it is a simplistic approach. So as a second estimate, we calculate the expected signal-to-noise at the end of the mission\footnote{Estimated until June 20, 2019} $S/N_{Nmeas}$ (see Table \ref{GaiaPar}) using the number of measurements provided by the \emph{Gaia} Observation Forecast Tool\footnote{http://gaia.esac.esa.int/gost/}. If we combine astrometric data with RVs, an astrometric orbit could be retrieved with an astrometric signal-to-noise down to 6.2 \citep{Sahlmann2011}. By the end of the mission, all three planets could be detected using a combined analysis with estimated values of $S/N_{Nmeas}=8.0$, $19.5$, and $69.2$ for HD17674, HD42012, and HD29021, respectively.\\

\begin{table*}[ht]
\caption{Data from \emph{Gaia} DR1 and estimated values. The number of \emph{Gaia} observations is determined for each target using the \emph{Gaia} Observation Forecast Tool. The S/N is calculated for one measurement, the measurements available in DR1, and the expected values at the end of the mission.}              
\label{GaiaPar}      
\centering                                      
	\begin{tabular}{l c c c c  c c c}          
	\hline\hline                        
	Target & $\alpha$ & Nmeas & Nmeas & 	S/N$_{1meas}$ & S/N$_{Nmeas}$ & Astrometric Excess Noise & $\Delta Q$\\
	Name &  [$\mu as$] & DR1 & End of Mission &   & & [mas] & \\ 
	\hline
	HD17674 & 27 & 32$^{(1)}$ & 101 & 0.8 & 8.0 & 0.33  & 3.04\\
	HD42012 & 83 & 5$^{(1)}$ & 66 & 2.4 & 19.5 & 0.28  & -\\
	HD29021 & 201$^{(2)}$ & 14$^{(1)}$ & 133 & 6.0 &69.2 & 0.52 & 2.15 \\       
	\end{tabular}	
	\begin{flushleft}
		\begin{small}
$^{(1)}$ Number of matched observations from the \emph{Gaia} archive. Using the Forecast Tool, the expected number of observations were 35, 5 and 19 respectively.\\
$^{(2)}$ The detectability of eccentric orbits will vary over the orbital phase, so the astrometric signal given for HD29021 corresponds to apastron.\\
\end{small}
\end{flushleft}
\end{table*}

The \emph{Gaia} DR1 released in September 2016 \citep{Lindegren2016} contains provisional astrometric results for over one billion sources brighter than G magnitude 20.7 for the first 418 days of the mission. Using DR1, we discuss a third indicator. 
An orbiting companion of planetary mass will produce a perturbation in the stellar motion, and this will be reflected as a deviation of the astrometric data from the five-parameter model. In \emph{Gaia} DR1, this value is called \emph{astrometric excess noise} $\epsilon_i$ [mas] and is listed in Table \ref{GaiaPar}. As \cite{Lindegren2016} describe in their work, $\epsilon_i=0$ when the source is astrometrically well behaved, and a value larger than 0 indicates that the residuals are statistically larger than expected. But since DR1 only presents a preliminary solution, the calibration modelling errors are high and this leads to a significant value ($\sim$ 0.5 mas) of the \emph{astrometric excess noise} for nearly all sources. Nevertheless, \cite{Lindegren2016} also mention that a source with a value of $\epsilon_i$ above 1-2 mas could indicate the presence of an astrometric binary or a problematic source. This is not the case for any of the $\epsilon_i$ listed in Table \ref{GaiaPar}. This is especially important for HD29021 b where we could not set an upper-mass limit from the Hipparcos data.\\

A final indicator worth mentioning is the \emph{astrometric} $\Delta Q$. This quantity is defined for the \emph{Tycho}-\emph{Gaia} (TGAS) astrometric solution and it is sensitive to the difference of Hipparcos and TGAS proper motions. It was first introduced by \cite{Michalik2014}, in the context of the Hundred Thousand Proper Motions project, and was also used by \cite{Lindegren2016} in their work on \emph{Gaia} DR1 with a slightly different definition of the parameter that only considers the differences on proper motions\\
For single stars, $\Delta Q$ is expected to follow a chi-squared distribution with two degrees of freedom \citep{Lindegren2016}. Deviations from this theoretical distribution could be caused, for example, by binaries with periods of 10-50 years that will have increased values of $\Delta Q$ \citep{Michalik2014}. This latter work also demonstrated a strong dependency of $\Delta Q$ on the quality of the Gaia solution, so even though the full sensitivity of this parameter is not reached in DR1, it will increase in the next Gaia releases. We compared the $\Delta Q$ values of the present TGAS solution for our targets (see Table \ref{GaiaPar}) with Fig. C.3 of \cite{Lindegren2016}. Our values are not significantly high and are placed in the region where the highest relative frequency is expected, near the theoretically expected distribution.

\section{Discussion and conclusions}
We have reported the detection of three new Jupiter-mass companions orbiting the solar-type stars HD17674, HD29021, and HD42012. We found no evidence of additional giant companions. As a reference for this population of objects, as of October 2016, the exoplanets.eu database \citep{exoplaneteu} listed around 200 planets with masses between 0.8 and 15 M$_{Jup}$ and periods longer than 600 days.\\
 The three host stars present a subsolar metallicity, which is unusual for stars with giant planetary companions \citep{Johnson2010}, but also reinforces the result found by \cite{Adibekyan2013}, who showed that planets orbiting metal-poor stars have longer periods than those around metal-rich ones. The fact that these companions all make, as far as our analysis can tell, single giant-planet systems, may also be linked to the low heavy-element content of their birth nebulosity.\\
We expect that for the last \emph{Gaia} data release (programmed for 2022) these three new planets will be fully characterized with a joint astrometric and radial velocity analysis. 

\begin{acknowledgements}
We gratefully acknowledge the Programme National de Plan\'etologie (telescope time attribution and financial support) of CNRS/INSU, the Swiss National Science Foundation, and the Agence Nationale de la Recherche (grant ANR-08-JCJC-0102-01) for their support. We warmly thank the OHP staff for their support on the 1.93 m telescope. J.R. acknowledges support from CONICYT-Becas Chile (grant 72140583). J.S. is supported by an ESA Research Fellowship in Space Science. A.S. is supported by the European Union under a Marie Curie Intra-European Fellowship for Career Development with reference FP7-PEOPLE-2013-IEF, number 627202. This work has been carried out in the frame of the National Centre for Competence in Research “PlanetS” supported by the Swiss National Science Foundation (SNSF). D.S., R.F.D., N.A., V.B., D.E., F.P., and S.U. acknowledge the financial support of the SNSF. P.A.W. acknowledges the support of the French Agence Nationale de la Recherche (ANR), under programme ANR-12-BS05-0012 “Exo-Atmos”. The IA team was supported by Funda\c{c}\~ao para a Ci\^encia e a Tecnologia (FCT) (project ref. PTDC/FIS-AST/1526/2014) through national funds and by FEDER through COMPETE2020 (ref. POCI-01-0145-FEDER-016886), and through grant UID/FIS/04434/2013 (POCI-01-0145-FEDER-007672).
N.C.S. was supported by FCT through the Investigador FCT contract reference IF/00169/2012 and POPH/FSE (EC) by FEDER funding through the programme “Programa Operacional de Factores de Competitividade - COMPETE”. This research has made use of the SIMBAD database and of the VizieR catalogue access tool operated at CDS, France.
This work has made use of data from the European Space Agency (ESA) mission {\it Gaia} (\url{http://www.cosmos.esa.int/gaia}), processed by the {\it Gaia} Data Processing and Analysis Consortium (DPAC, \url{http://www.cosmos.esa.int/web/gaia/dpac/consortium}). Funding for the DPAC has been provided by national institutions, in particular the institutions participating in the {\it Gaia} Multilateral Agreement.
\end{acknowledgements}

%
%

\bibliographystyle{aa} 
\bibliography{RPE_1} 

\begin{thebibliography}{54}
\expandafter\ifx\csname natexlab\endcsname\relax\def\natexlab#1{#1}\fi

\bibitem[{{Adibekyan} {et~al.}(2013){Adibekyan}, {Figueira}, {Santos},
  {Mortier}, {Mordasini}, {Delgado Mena}, {Sousa}, {Correia}, {Israelian}, \&
  {Oshagh}}]{Adibekyan2013}
{Adibekyan}, V.~Z., {Figueira}, P., {Santos}, N.~C., {et~al.} 2013, \aap, 560,
  A51

\bibitem[{{Baranne} {et~al.}(1996){Baranne}, {Queloz}, {Mayor}, {Adrianzyk},
  {Knispel}, {Kohler}, {Lacroix}, {Meunier}, {Rimbaud}, \& {Vin}}]{Baranne1996}
{Baranne}, A., {Queloz}, D., {Mayor}, M., {et~al.} 1996, \aaps, 119, 373

\bibitem[{{Benedict} {et~al.}(2010){Benedict}, {McArthur}, {Bean}, {Barnes},
  {Harrison}, {Hatzes}, {Martioli}, \& {Nelan}}]{Benedict2010}
{Benedict}, G.~F., {McArthur}, B.~E., {Bean}, J.~L., {et~al.} 2010, \aj, 139,
  1844

\bibitem[{{Boisse} {et~al.}(2011){Boisse}, {Bouchy}, {H{\'e}brard}, {Bonfils},
  {Santos}, \& {Vauclair}}]{Boisse2011a}
{Boisse}, I., {Bouchy}, F., {H{\'e}brard}, G., {et~al.} 2011, \aap, 528, A4

\bibitem[{{Boisse} {et~al.}(2010){Boisse}, {Eggenberger}, {Santos}, {Lovis},
  {Bouchy}, {H{\'e}brard}, {Arnold}, {Bonfils}, {Delfosse}, {Desort},
  {D{\'{\i}}az}, {Ehrenreich}, {Forveille}, {Gallenne}, {Lagrange}, {Moutou},
  {Udry}, {Pepe}, {Perrier}, {Perruchot}, {Pont}, {Queloz}, {Santerne},
  {S{\'e}gransan}, \& {Vidal-Madjar}}]{Boisse2010}
{Boisse}, I., {Eggenberger}, A., {Santos}, N.~C., {et~al.} 2010, \aap, 523, A88

\bibitem[{{Boisse} {et~al.}(2012){Boisse}, {Pepe}, {Perrier}, {Queloz},
  {Bonfils}, {Bouchy}, {Santos}, {Arnold}, {Beuzit}, {D{\'{\i}}az}, {Delfosse},
  {Eggenberger}, {Ehrenreich}, {Forveille}, {H{\'e}brard}, {Lagrange}, {Lovis},
  {Mayor}, {Moutou}, {Naef}, {Santerne}, {S{\'e}gransan}, {Sivan}, \&
  {Udry}}]{SOPHIEV}
{Boisse}, I., {Pepe}, F., {Perrier}, C., {et~al.} 2012, \aap, 545, A55

\bibitem[{{Bonomo} {et~al.}(2010){Bonomo}, {Santerne}, {Alonso}, {Gazzano},
  {Havel}, {Aigrain}, {Auvergne}, {Baglin}, {Barbieri}, {Barge}, {Benz},
  {Bord{\'e}}, {Bouchy}, {Bruntt}, {Cabrera}, {Collier Cameron}, {Carone},
  {Carpano}, {Csizmadia}, {Deleuil}, {Deeg}, {Dvorak}, {Erikson},
  {Ferraz-Mello}, {Fridlund}, {Gandolfi}, {Gillon}, {Guenther}, {Guillot},
  {Hatzes}, {H{\'e}brard}, {Jorda}, {Lammer}, {Lanza}, {L{\'e}ger}, {Llebaria},
  {Mayor}, {Mazeh}, {Moutou}, {Ollivier}, {P{\"a}tzold}, {Pepe}, {Queloz},
  {Rauer}, {Rouan}, {Samuel}, {Schneider}, {Tingley}, {Udry}, \&
  {Wuchterl}}]{Bonomo2010}
{Bonomo}, A.~S., {Santerne}, A., {Alonso}, R., {et~al.} 2010, \aap, 520, A65

\bibitem[{{Bouchy} {et~al.}(2013){Bouchy}, {D{\'{\i}}az}, {H{\'e}brard},
  {Arnold}, {Boisse}, {Delfosse}, {Perruchot}, \& {Santerne}}]{Bouchy2013}
{Bouchy}, F., {D{\'{\i}}az}, R.~F., {H{\'e}brard}, G., {et~al.} 2013, \aap,
  549, A49

\bibitem[{{Bouchy} {et~al.}(2009{\natexlab{a}}){Bouchy}, {H{\'e}brard}, {Udry},
  {Delfosse}, {Boisse}, {Desort}, {Bonfils}, {Eggenberger}, {Ehrenreich},
  {Forveille}, {Lagrange}, {Le Coroller}, {Lovis}, {Moutou}, {Pepe}, {Perrier},
  {Pont}, {Queloz}, {Santos}, {S{\'e}gransan}, \& {Vidal-Madjar}}]{Bouchy2009a}
{Bouchy}, F., {H{\'e}brard}, G., {Udry}, S., {et~al.} 2009{\natexlab{a}}, \aap,
  505, 853

\bibitem[{{Bouchy} {et~al.}(2009{\natexlab{b}}){Bouchy}, {Isambert}, {Lovis},
  {Boisse}, {Figueira}, {H{\'e}brard}, \& {Pepe}}]{Bouchy2009cti}
{Bouchy}, F., {Isambert}, J., {Lovis}, C., {et~al.} 2009{\natexlab{b}}, in EAS
  Publications Series, Vol.~37, EAS Publications Series, ed. P.~{Kern},
  247--253

\bibitem[{{Bouchy} {et~al.}(2016){Bouchy}, {S{\'e}gransan}, {D{\'{\i}}az},
  {Forveille}, {Boisse}, {Arnold}, {Astudillo-Defru}, {Beuzit}, {Bonfils},
  {Borgniet}, {Bourrier}, {Courcol}, {Delfosse}, {Demangeon}, {Delorme},
  {Ehrenreich}, {H{\'e}brard}, {Lagrange}, {Mayor}, {Montagnier}, {Moutou},
  {Naef}, {Pepe}, {Perrier}, {Queloz}, {Rey}, {Sahlmann}, {Santerne}, {Santos},
  {Sivan}, {Udry}, \& {Wilson}}]{Bouchy2016}
{Bouchy}, F., {S{\'e}gransan}, D., {D{\'{\i}}az}, R.~F., {et~al.} 2016, \aap,
  585, A46

\bibitem[{{Bressan} {et~al.}(2012){Bressan}, {Marigo}, {Girardi}, {Salasnich},
  {Dal Cero}, {Rubele}, \& {Nanni}}]{Bressan2012}
{Bressan}, A., {Marigo}, P., {Girardi}, L., {et~al.} 2012, \mnras, 427, 127

\bibitem[{{Casertano} {et~al.}(2008){Casertano}, {Lattanzi}, {Sozzetti},
  {Spagna}, {Jancart}, {Morbidelli}, {Pannunzio}, {Pourbaix}, \&
  {Queloz}}]{Casertano2008}
{Casertano}, S., {Lattanzi}, M.~G., {Sozzetti}, A., {et~al.} 2008, \aap, 482,
  699

\bibitem[{{Cincunegui} {et~al.}(2007{\natexlab{a}}){Cincunegui}, {D{\'{\i}}az},
  \& {Mauas}}]{Cincunegui2007a}
{Cincunegui}, C., {D{\'{\i}}az}, R.~F., \& {Mauas}, P.~J.~D.
  2007{\natexlab{a}}, \aap, 461, 1107

\bibitem[{{Cincunegui} {et~al.}(2007{\natexlab{b}}){Cincunegui}, {D{\'{\i}}az},
  \& {Mauas}}]{Cincunegui2007b}
{Cincunegui}, C., {D{\'{\i}}az}, R.~F., \& {Mauas}, P.~J.~D.
  2007{\natexlab{b}}, \aap, 469, 309

\bibitem[{{Correia} {et~al.}(2010){Correia}, {Couetdic}, {Laskar}, {Bonfils},
  {Mayor}, {Bertaux}, {Bouchy}, {Delfosse}, {Forveille}, {Lovis}, {Pepe},
  {Perrier}, {Queloz}, \& {Udry}}]{Correia2010}
{Correia}, A.~C.~M., {Couetdic}, J., {Laskar}, J., {et~al.} 2010, \aap, 511,
  A21

\bibitem[{{Courcol} {et~al.}(2015){Courcol}, {Bouchy}, {Pepe}, {Santerne},
  {Delfosse}, {Arnold}, {Astudillo-Defru}, {Boisse}, {Bonfils}, {Borgniet},
  {Bourrier}, {Cabrera}, {Deleuil}, {Demangeon}, {D{\'{\i}}az}, {Ehrenreich},
  {Forveille}, {H{\'e}brard}, {Lagrange}, {Montagnier}, {Moutou}, {Rey},
  {Santos}, {S{\'e}gransan}, {Udry}, \& {Wilson}}]{Courcol2015}
{Courcol}, B., {Bouchy}, F., {Pepe}, F., {et~al.} 2015, \aap, 581, A38

\bibitem[{{da Silva} {et~al.}(2006){da Silva}, {Girardi}, {Pasquini},
  {Setiawan}, {von der L{\"u}he}, {de Medeiros}, {Hatzes}, {D{\"o}llinger}, \&
  {Weiss}}]{daSilva2006}
{da Silva}, L., {Girardi}, L., {Pasquini}, L., {et~al.} 2006, \aap, 458, 609

\bibitem[{{Delisle} {et~al.}(2016){Delisle}, {S{\'e}gransan}, {Buchschacher},
  \& {Alesina}}]{Delisle2016}
{Delisle}, J.-B., {S{\'e}gransan}, D., {Buchschacher}, N., \& {Alesina}, F.
  2016, \aap, 590, A134

\bibitem[{{D{\'{\i}}az} {et~al.}(2014){D{\'{\i}}az}, {Almenara}, {Santerne},
  {Moutou}, {Lethuillier}, \& {Deleuil}}]{Diaz2014}
{D{\'{\i}}az}, R.~F., {Almenara}, J.~M., {Santerne}, A., {et~al.} 2014, \mnras,
  441, 983

\bibitem[{{D{\'{\i}}az} {et~al.}(2016{\natexlab{a}}){D{\'{\i}}az}, {Rey},
  {Demangeon}, {H{\'e}brard}, {Boisse}, {Arnold}, {Astudillo-Defru}, {Beuzit},
  {Bonfils}, {Borgniet}, {Bouchy}, {Bourrier}, {Courcol}, {Deleuil},
  {Delfosse}, {Ehrenreich}, {Forveille}, {Lagrange}, {Mayor}, {Moutou}, {Pepe},
  {Queloz}, {Santerne}, {Santos}, {Sahlmann}, {S{\'e}gransan}, {Udry}, \&
  {Wilson}}]{Diaz2016}
{D{\'{\i}}az}, R.~F., {Rey}, J., {Demangeon}, O., {et~al.} 2016{\natexlab{a}},
  \aap, 591, A146

\bibitem[{{D{\'{\i}}az} {et~al.}(2012){D{\'{\i}}az}, {Santerne}, {Sahlmann},
  {H{\'e}brard}, {Eggenberger}, {Santos}, {Moutou}, {Arnold}, {Boisse},
  {Bonfils}, {Bouchy}, {Delfosse}, {Desort}, {Ehrenreich}, {Forveille},
  {Lagrange}, {Lovis}, {Pepe}, {Perrier}, {Queloz}, {S{\'e}gransan}, {Udry}, \&
  {Vidal-Madjar}}]{Diaz2012}
{D{\'{\i}}az}, R.~F., {Santerne}, A., {Sahlmann}, J., {et~al.} 2012, \aap, 538,
  A113

\bibitem[{{D{\'{\i}}az} {et~al.}(2016{\natexlab{b}}){D{\'{\i}}az},
  {S{\'e}gransan}, {Udry}, {Lovis}, {Pepe}, {Dumusque}, {Marmier}, {Alonso},
  {Benz}, {Bouchy}, {Coffinet}, {Collier Cameron}, {Deleuil}, {Figueira},
  {Gillon}, {Lo Curto}, {Mayor}, {Mordasini}, {Motalebi}, {Moutou}, {Pollacco},
  {Pompei}, {Queloz}, {Santos}, \& {Wyttenbach}}]{Diaz2016harps}
{D{\'{\i}}az}, R.~F., {S{\'e}gransan}, D., {Udry}, S., {et~al.}
  2016{\natexlab{b}}, \aap, 585, A134

\bibitem[{{ESA}(1997)}]{Hipparcos1997}
{ESA}, ed. 1997, ESA Special Publication, Vol. 1200, {The HIPPARCOS and TYCHO
  catalogues. Astrometric and photometric star catalogues derived from the ESA
  HIPPARCOS Space Astrometry Mission}

\bibitem[{{Gaia Collaboration}(2016)}]{GaiaMission2016}
{Gaia Collaboration}. 2016, ArXiv e-prints [\eprint[arXiv]{1609.04153}]

\bibitem[{{Gaia Collaboration} {et~al.}(2016){Gaia Collaboration}, {Brown},
  {Vallenari}, {Prusti}, {de Bruijne}, {Mignard}, {Drimmel}, \&
  {co-authors}}]{GaiaDR12016}
{Gaia Collaboration}, {Brown}, A.~G.~A., {Vallenari}, A., {et~al.} 2016, ArXiv
  e-prints [\eprint[arXiv]{1609.04172}]

\bibitem[{{Gomes da Silva} {et~al.}(2014){Gomes da Silva}, {Santos}, {Boisse},
  {Dumusque}, \& {Lovis}}]{Gomesdasilva2014}
{Gomes da Silva}, J., {Santos}, N.~C., {Boisse}, I., {Dumusque}, X., \&
  {Lovis}, C. 2014, \aap, 566, A66

\bibitem[{{H{\'e}brard} {et~al.}(2016){H{\'e}brard}, {Arnold}, {Forveille},
  {Correia}, {Laskar}, {Bonfils}, {Boisse}, {D{\'{\i}}az}, {Hagelberg},
  {Sahlmann}, {Santos}, {Astudillo-Defru}, {Borgniet}, {Bouchy}, {Bourrier},
  {Courcol}, {Delfosse}, {Deleuil}, {Demangeon}, {Ehrenreich}, {Gregorio},
  {Jovanovic}, {Labrevoir}, {Lagrange}, {Lovis}, {Lozi}, {Moutou},
  {Montagnier}, {Pepe}, {Rey}, {Santerne}, {S{\'e}gransan}, {Udry},
  {Vanhuysse}, {Vigan}, \& {Wilson}}]{Hebrard2016}
{H{\'e}brard}, G., {Arnold}, L., {Forveille}, T., {et~al.} 2016, \aap, 588,
  A145

\bibitem[{{Johnson} {et~al.}(2010){Johnson}, {Aller}, {Howard}, \&
  {Crepp}}]{Johnson2010}
{Johnson}, J.~A., {Aller}, K.~M., {Howard}, A.~W., \& {Crepp}, J.~R. 2010,
  \pasp, 122, 905

\bibitem[{{Kopparapu} {et~al.}(2013){Kopparapu}, {Ramirez}, {Kasting}, {Eymet},
  {Robinson}, {Mahadevan}, {Terrien}, {Domagal-Goldman}, {Meadows}, \&
  {Deshpande}}]{Kopparapu2013}
{Kopparapu}, R.~K., {Ramirez}, R., {Kasting}, J.~F., {et~al.} 2013, \apj, 765,
  131

\bibitem[{{Lindegren} {et~al.}(2016){Lindegren}, {Lammers}, {Bastian},
  {Hern{\'a}ndez}, {Klioner}, {Hobbs}, {Bombrun}, {Michalik}, {Ramos-Lerate},
  {Butkevich}, {Comoretto}, {Joliet}, {Holl}, {Hutton}, {Parsons},
  {Steidelm{\"u}ller}, {Abbas}, {Altmann}, {Andrei}, {Anton}, {Bach},
  {Barache}, {Becciani}, {Berthier}, {Bianchi}, {Biermann}, {Bouquillon},
  {Bourda}, {Br{\"u}semeister}, {Bucciarelli}, {Busonero}, {Carlucci},
  {Casta{\~n}eda}, {Charlot}, {Clotet}, {Crosta}, {Davidson}, {de Felice},
  {Drimmel}, {Fabricius}, {Fienga}, {Figueras}, {Fraile}, {Gai}, {Garralda},
  {Geyer}, {Gonz{\'a}lez-Vidal}, {Guerra}, {Hambly}, {Hauser}, {Jordan},
  {Lattanzi}, {Lenhardt}, {Liao}, {L{\"o}ffler}, {McMillan}, {Mignard}, {Mora},
  {Morbidelli}, {Portell}, {Riva}, {Sarasso}, {Serraller}, {Siddiqui}, {Smart},
  {Spagna}, {Stampa}, {Steele}, {Taris}, {Torra}, {van Reeven}, {Vecchiato},
  {Zschocke}, {de Bruijne}, {Gracia}, {Raison}, {Lister}, {Marchant},
  {Messineo}, {Soffel}, {Osorio}, {de Torres}, \& {O'Mullane}}]{Lindegren2016}
{Lindegren}, L., {Lammers}, U., {Bastian}, U., {et~al.} 2016, ArXiv e-prints
  [\eprint[arXiv]{1609.04303}]

\bibitem[{{Mayor} \& {Queloz}(1995)}]{Mayor1995}
{Mayor}, M. \& {Queloz}, D. 1995, \nat, 378, 355

\bibitem[{{Michalik} {et~al.}(2014){Michalik}, {Lindegren}, {Hobbs}, \&
  {Lammers}}]{Michalik2014}
{Michalik}, D., {Lindegren}, L., {Hobbs}, D., \& {Lammers}, U. 2014, \aap, 571,
  A85

\bibitem[{{Montes} {et~al.}(1995){Montes}, {Fernandez-Figueroa}, {de Castro},
  \& {Cornide}}]{Montes1995}
{Montes}, D., {Fernandez-Figueroa}, M.~J., {de Castro}, E., \& {Cornide}, M.
  1995, \aap, 294, 165

\bibitem[{{Muterspaugh} {et~al.}(2010){Muterspaugh}, {Lane}, {Kulkarni},
  {Konacki}, {Burke}, {Colavita}, {Shao}, {Hartkopf}, {Boss}, \&
  {Williamson}}]{Muterspaugh2010}
{Muterspaugh}, M.~W., {Lane}, B.~F., {Kulkarni}, S.~R., {et~al.} 2010, \aj,
  140, 1657

\bibitem[{{Neveu-VanMalle} {et~al.}(2016){Neveu-VanMalle}, {Queloz},
  {Anderson}, {Brown}, {Collier Cameron}, {Delrez}, {D{\'{\i}}az}, {Gillon},
  {Hellier}, {Jehin}, {Lister}, {Pepe}, {Rojo}, {S{\'e}gransan}, {Triaud},
  {Turner}, \& {Udry}}]{Neveu2016}
{Neveu-VanMalle}, M., {Queloz}, D., {Anderson}, D.~R., {et~al.} 2016, \aap,
  586, A93

\bibitem[{{Noyes} {et~al.}(1984){Noyes}, {Hartmann}, {Baliunas}, {Duncan}, \&
  {Vaughan}}]{Noyes1984}
{Noyes}, R.~W., {Hartmann}, L.~W., {Baliunas}, S.~L., {Duncan}, D.~K., \&
  {Vaughan}, A.~H. 1984, \apj, 279, 763

\bibitem[{{Pasquini} \& {Pallavicini}(1991)}]{Pasquini1991}
{Pasquini}, L. \& {Pallavicini}, R. 1991, \aap, 251, 199

\bibitem[{{Pepe} {et~al.}(2002){Pepe}, {Mayor}, {Galland}, {Naef}, {Queloz},
  {Santos}, {Udry}, \& {Burnet}}]{Pepe2002}
{Pepe}, F., {Mayor}, M., {Galland}, F., {et~al.} 2002, \aap, 388, 632

\bibitem[{{Perruchot} {et~al.}(2008){Perruchot}, {Kohler}, {Bouchy}, {Richaud},
  {Richaud}, {Moreaux}, {Merzougui}, {Sottile}, {Hill}, {Knispel}, {Regal},
  {Meunier}, {Ilovaisky}, {Le Coroller}, {Gillet}, {Schmitt}, {Pepe}, {Fleury},
  {Sosnowska}, {Vors}, {M{\'e}gevand}, {Blanc}, {Carol}, {Point}, {Laloge}, \&
  {Brunel}}]{Perruchot2008}
{Perruchot}, S., {Kohler}, D., {Bouchy}, F., {et~al.} 2008, in \procspie, Vol.
  7014, Ground-based and Airborne Instrumentation for Astronomy II, 70140J

\bibitem[{{Perryman} {et~al.}(2014){Perryman}, {Hartman}, {Bakos}, \&
  {Lindegren}}]{Perryman2014}
{Perryman}, M., {Hartman}, J., {Bakos}, G.~{\'A}., \& {Lindegren}, L. 2014,
  \apj, 797, 14

\bibitem[{{Queloz} {et~al.}(2001){Queloz}, {Henry}, {Sivan}, {Baliunas},
  {Beuzit}, {Donahue}, {Mayor}, {Naef}, {Perrier}, \& {Udry}}]{Queloz2001}
{Queloz}, D., {Henry}, G.~W., {Sivan}, J.~P., {et~al.} 2001, \aap, 379, 279

\bibitem[{{Robertson} {et~al.}(2013){Robertson}, {Endl}, {Cochran}, \&
  {Dodson-Robinson}}]{Robertson2013}
{Robertson}, P., {Endl}, M., {Cochran}, W.~D., \& {Dodson-Robinson}, S.~E.
  2013, \apj, 764, 3

\bibitem[{{Robertson} {et~al.}(2014){Robertson}, {Mahadevan}, {Endl}, \&
  {Roy}}]{Robertson2014}
{Robertson}, P., {Mahadevan}, S., {Endl}, M., \& {Roy}, A. 2014, Science, 345,
  440

\bibitem[{{Sahlmann} {et~al.}(2016){Sahlmann}, {Lazorenko}, {S{\'e}gransan},
  {Astudillo-Defru}, {Bonfils}, {Delfosse}, {Forveille}, {Hagelberg}, {Lo
  Curto}, {Pepe}, {Queloz}, {Udry}, \& {Zimmerman}}]{Sahlmann2016}
{Sahlmann}, J., {Lazorenko}, P.~F., {S{\'e}gransan}, D., {et~al.} 2016, ArXiv
  e-prints [\eprint[arXiv]{1608.00918}]

\bibitem[{{Sahlmann} {et~al.}(2011{\natexlab{a}}){Sahlmann}, {Lovis}, {Queloz},
  \& {S{\'e}gransan}}]{Sahlmann2011}
{Sahlmann}, J., {Lovis}, C., {Queloz}, D., \& {S{\'e}gransan}, D.
  2011{\natexlab{a}}, \aap, 528, L8

\bibitem[{{Sahlmann} {et~al.}(2011{\natexlab{b}}){Sahlmann}, {S{\'e}gransan},
  {Queloz}, {Udry}, {Santos}, {Marmier}, {Mayor}, {Naef}, {Pepe}, \&
  {Zucker}}]{Sahlmann2011b}
{Sahlmann}, J., {S{\'e}gransan}, D., {Queloz}, D., {et~al.} 2011{\natexlab{b}},
  \aap, 525, A95

\bibitem[{{Santerne} {et~al.}(2012){Santerne}, {D{\'{\i}}az}, {Moutou},
  {Bouchy}, {H{\'e}brard}, {Almenara}, {Bonomo}, {Deleuil}, \&
  {Santos}}]{Santerne2012}
{Santerne}, A., {D{\'{\i}}az}, R.~F., {Moutou}, C., {et~al.} 2012, \aap, 545,
  A76

\bibitem[{{Santos} {et~al.}(2004){Santos}, {Israelian}, \&
  {Mayor}}]{Santos2004}
{Santos}, N.~C., {Israelian}, G., \& {Mayor}, M. 2004, \aap, 415, 1153

\bibitem[{{Santos} {et~al.}(2013){Santos}, {Sousa}, {Mortier}, {Neves},
  {Adibekyan}, {Tsantaki}, {Delgado Mena}, {Bonfils}, {Israelian}, {Mayor}, \&
  {Udry}}]{Santos2013}
{Santos}, N.~C., {Sousa}, S.~G., {Mortier}, A., {et~al.} 2013, \aap, 556, A150

\bibitem[{{Schneider} {et~al.}(2011){Schneider}, {Dedieu}, {Le Sidaner},
  {Savalle}, \& {Zolotukhin}}]{exoplaneteu}
{Schneider}, J., {Dedieu}, C., {Le Sidaner}, P., {Savalle}, R., \&
  {Zolotukhin}, I. 2011, \aap, 532, A79

\bibitem[{{Sousa} {et~al.}(2008){Sousa}, {Santos}, {Mayor}, {Udry},
  {Casagrande}, {Israelian}, {Pepe}, {Queloz}, \& {Monteiro}}]{Sousa2008}
{Sousa}, S.~G., {Santos}, N.~C., {Mayor}, M., {et~al.} 2008, \aap, 487, 373

\bibitem[{{Torres} {et~al.}(2010){Torres}, {Andersen}, \&
  {Gim{\'e}nez}}]{Torres2010}
{Torres}, G., {Andersen}, J., \& {Gim{\'e}nez}, A. 2010, \aapr, 18, 67

\bibitem[{{van Leeuwen}(2007)}]{vanLeeuwen2007}
{van Leeuwen}, F. 2007, \aap, 474, 653

\end{thebibliography}

\longtab[1]{
\begin{longtable}{lccc}
\caption{Radial velocities of HD17674.}\\
\hline
\hline
	BJD & RV & $\sigma (RV)$ & Instrument\\
	$[-2400000$ $days]$ & [$km/s$] & [$km/s$] & \\
\hline
\endfirsthead
\caption{Continued.} \\
\hline
	BJD & RV & $\sigma (RV)$ & Instrument\\
	$[-2400000$ $days]$ & [$km/s$] & [$km/s$] & \\
\hline
\endhead
\hline
\endfoot
\hline
\endlastfoot
50731.5728	&10.547	&0.009	&ELODIE	\\
	50733.5803	&10.556	&0.012 	&ELODIE	\\
	51087.6079	&10.549	&0.009	&ELODIE	\\
	51423.6193	&10.539	&0.009	&ELODIE	\\
	51445.6047	&10.560	&0.008	&ELODIE	\\
	51560.3474	&10.567	&0.010	&ELODIE	\\
	51853.5015	&10.533	&0.010	&ELODIE	\\
	51952.3135	&10.528	&0.009	&ELODIE	\\
	54366.6249	&10.5851	&0.0062	&SOPHIE \\
	54367.6473	&10.5847	&0.0061	&SOPHIE \\
	54375.5863	&10.5722	&0.0061	&SOPHIE \\
	54379.5426	&10.5764	&0.0064	&SOPHIE \\
	54379.5452	&10.5798	&0.0064	&SOPHIE \\
	54379.5478	&10.5751	&0.0065	&SOPHIE \\
	54381.4875	&10.5785	&0.0062	&SOPHIE \\
	54689.6326	&10.6346	&0.0062	&SOPHIE \\
	54707.6312	&10.6378	&0.0061	&SOPHIE \\
	54852.3384	&10.6145	&0.0064	&SOPHIE \\
	54857.3196	&10.5975	&0.0064	&SOPHIE \\
	54881.2713	&10.6026	&0.0064	&SOPHIE \\
	54894.3095	&10.5980	&0.0065	&SOPHIE \\
	55050.6272	&10.5988	&0.0062	&SOPHIE \\
	55059.6243	&10.5951	&0.0062	&SOPHIE \\
	55061.6267	&10.5942	&0.0061	&SOPHIE \\
	55062.6359	&10.6009	&0.0062	&SOPHIE \\
	55068.6137	&10.6131	&0.0063	&SOPHIE \\
	55071.6209	&10.5940	&0.0062	&SOPHIE \\
	55075.6095	&10.5986	&0.0061	&SOPHIE \\
	55078.6028	&10.6075	&0.0063	&SOPHIE \\
	55140.5422	&10.6074	&0.0062	&SOPHIE \\
	55141.5476	&10.6154	&0.0062	&SOPHIE \\
	55148.4721	&10.6026	&0.0061	&SOPHIE \\
	55449.6038	&10.6226	&0.0062	&SOPHIE \\
	55450.5773	&10.6184	&0.0061	&SOPHIE \\
	55453.6185	&10.6160	&0.0062	&SOPHIE \\
	55479.5179	&10.6100	&0.0065	&SOPHIE \\
	55481.5293	&10.6059	&0.0061	&SOPHIE \\
	55483.5386	&10.6075	&0.0062	&SOPHIE \\
	55484.4508	&10.6097	&0.0062	&SOPHIE \\
	55495.4945	&10.6042	&0.0063	&SOPHIE \\
	55498.4640	&10.5938	&0.0061	&SOPHIE \\
	55527.4474	&10.5964	&0.0063	&SOPHIE \\
	55551.4847	&10.5887	&0.0062	&SOPHIE \\
	55559.2883	&10.5908	&0.0062	&SOPHIE \\
	55564.3481	&10.5927	&0.0062	&SOPHIE \\
	56197.6216	&10.5891	&0.0019	&SOPHIE+ \\
	56229.5802	&10.5864	&0.0018	&SOPHIE+ \\
	56230.6107	&10.5796	&0.0015	&SOPHIE+ \\
	56251.5219	&10.5885	&0.0014	&SOPHIE+ \\
	56252.4853	&10.5864	&0.0014	&SOPHIE+ \\
	56285.3763	&10.5901	&0.0014	&SOPHIE+ \\
	56290.4409	&10.5920	&0.0017	&SOPHIE+ \\
	56291.4018	&10.5943	&0.0017	&SOPHIE+ \\
	56292.4112	&10.5925	&0.0019	&SOPHIE+ \\
	56296.4632	&10.5872	&0.0016	&SOPHIE+ \\
	56318.2903	&10.5895	&0.0020	&SOPHIE+ \\
	56321.3202	&10.5847	&0.0019	&SOPHIE+ \\
	56625.4795	&10.6281	&0.0017	&SOPHIE+ \\
	56626.3869	&10.6248	&0.0017	&SOPHIE+ \\
	56627.3911	&10.6235	&0.0016	&SOPHIE+ \\
	56628.3784	&10.6238	&0.0016	&SOPHIE+ \\
	56629.3792	&10.6189	&0.0015	&SOPHIE+ \\
	56630.3612	&10.6244	&0.0014	&SOPHIE+ \\
	56631.3437	&10.6227	&0.0014	&SOPHIE+ \\
	56641.3450	&10.6249	&0.0016	&SOPHIE+ \\
	56642.4651	&10.6200	&0.0016	&SOPHIE+ \\
	56643.4579	&10.6225	&0.0014	&SOPHIE+ \\
	56654.3328	&10.6190	&0.0017	&SOPHIE+ \\
	56655.4128	&10.6171	&0.0024	&SOPHIE+ \\
	56656.3454	&10.6220	&0.0020	&SOPHIE+ \\
	56657.3921	&10.6240	&0.0022	&SOPHIE+ \\
	56682.3844	&10.6150	&0.0023	&SOPHIE+ \\
	56683.2899	&10.6168	&0.0019	&SOPHIE+ \\
	57079.2841	&10.6147	&0.0018	&SOPHIE+ \\
	57081.3406	&10.6083	&0.0022	&SOPHIE+ \\
	57082.2718	&10.6197	&0.0014	&SOPHIE+ \\
	57084.3132	&10.6177	&0.0016	&SOPHIE+ \\
	57088.3042	&10.6187	&0.0018	&SOPHIE+ \\
	57090.2717	&10.6217	&0.0019	&SOPHIE+ \\
	57270.5852	&10.6170	&0.0019	&SOPHIE+ \\
	57283.5606	&10.6186	&0.0017	&SOPHIE+ \\
	57284.5766	&10.6245	&0.0017	&SOPHIE+ \\
	57289.6718	&10.6187	&0.0018	&SOPHIE+ \\
	57318.5297	&10.6091	&0.0015	&SOPHIE+ \\
	57319.5255	&10.6081	&0.0014	&SOPHIE+ \\
	57326.5356	&10.6054	&0.0013	&SOPHIE+ \\
	57327.5446	&10.6091	&0.0014	&SOPHIE+ \\
	57328.4815	&10.6038	&0.0014	&SOPHIE+ \\
	57342.5121	&10.6090	&0.0013	&SOPHIE+ \\
	57345.4750	&10.6046	&0.0015	&SOPHIE+ \\
	57349.4221	&10.5970	&0.0018	&SOPHIE+ \\
	57373.4358	&10.6005	&0.0013	&SOPHIE+ \\
	57375.4532	&10.6020	&0.0016	&SOPHIE+ \\
	57383.3395	&10.5978	&0.0014	&SOPHIE+ \\
	57384.3697	&10.6004	&0.0014	&SOPHIE+ \\
	57385.4349	&10.5973	&0.0016	&SOPHIE+ \\
	57393.4056	&10.5952	&0.0016	&SOPHIE+ \\
	57394.3422	&10.5982	&0.0018	&SOPHIE+ \\
	57404.3579	&10.5958	&0.0020	&SOPHIE+ \\
	57412.2578	&10.5954	&0.0013	&SOPHIE+ \\
	57440.3545	&10.5943	&0.0029	&SOPHIE+ \\                  
\end{longtable}
}

   \longtab[1]{
\begin{longtable}{lccc}
\caption{Radial velocities of HD42012.}\\
\hline
\hline
	BJD & RV & $\sigma (RV)$ & Instrument\\
	$[-2400000$ $days]$ & [$km/s$] & [$km/s$] & \\
\hline
\endfirsthead
\caption{Continued.} \\
\hline
	BJD & RV & $\sigma (RV)$ & Instrument\\
	$[-2400000$ $days]$ & [$km/s$] & [$km/s$] & \\
\hline
\endhead
\hline
\endfoot
\hline
\endlastfoot
53066.3767	&41.3738	&0.0150	&ELODIE \\
54461.5393	&41.4771	&0.0067	&SOPHIE \\
54463.5387	&41.4559	&0.0061 &SOPHIE \\
55852.5515	&41.3813	&0.0034 &SOPHIE+\\
55879.5723	&41.3899	&0.0037 &SOPHIE+\\
56289.5602	&41.4430	&0.0036 &SOPHIE+\\
56326.3670	&41.4211	&0.0044 &SOPHIE+\\
56345.3569	&41.4359	&0.0035 &SOPHIE+\\
56354.3659	&41.4273	&0.0034 &SOPHIE+\\
56587.6472	&41.3755	&0.0046 &SOPHIE+\\
56610.6909	&41.3484	&0.0088 &SOPHIE+\\
56624.5463	&41.3834	&0.0035 &SOPHIE+\\
56625.5848	&41.3855	&0.0036 &SOPHIE+\\
56627.6562	&41.3734	&0.0053 &SOPHIE+\\
56628.6200	&41.3749	&0.0059 &SOPHIE+\\
56629.4549	&41.3682	&0.0036 &SOPHIE+\\
56630.4946	&41.3860	&0.0035 &SOPHIE+\\
56631.5222	&41.3816	&0.0035 &SOPHIE+\\
56663.4586	&41.3798	&0.0035 &SOPHIE+\\
56694.3507	&41.3676	&0.0074 &SOPHIE+\\
56734.4235	&41.3886	&0.0034 &SOPHIE+\\
56990.5750	&41.4444	&0.0040 &SOPHIE+\\
56999.5758	&41.4502	&0.0032 &SOPHIE+\\
57066.4235	&41.4574	&0.0032 &SOPHIE+\\
57076.3480	&41.4563	&0.0033 &SOPHIE+\\
57113.3182	&41.4525	&0.0046 &SOPHIE+\\
57152.3236	&41.4284	&0.0044 &SOPHIE+\\
57312.6930	&41.3994	&0.0033 &SOPHIE+\\
57326.7115	&41.3927	&0.0036 &SOPHIE+\\
57328.6237	&41.3960	&0.0033 &SOPHIE+\\
57394.5206	&41.3750	&0.0036 &SOPHIE+\\
57469.3745	&41.3798	&0.0035 &SOPHIE+\\
\end{longtable}
}

      \longtab[1]{
\begin{longtable}{lccc}
\caption{Radial velocities of HD29021.}\\
\hline
\hline
	BJD & RV & $\sigma (RV)$ & Instrument\\
	$[-2400000$ $days]$ & [$km/s$] & [$km/s$] & \\
\hline
\endfirsthead
\caption{Continued.} \\
\hline
	BJD & RV & $\sigma (RV)$ & Instrument\\
	$[-2400000$ $days]$ & [$km/s$] & [$km/s$] & \\
\hline
\endhead
\hline
\endfoot
\hline
\endlastfoot
55828.6566	&0.4326		&0.0036	&SOPHIE+\\
55837.6068	&0.4341		&0.0026	&SOPHIE+\\
55840.6330	&0.4273		&0.0028	&SOPHIE+\\
55843.6102	&0.4410		&0.0037	&SOPHIE+\\
55850.5553	&0.4339		&0.0041	&SOPHIE+\\
55851.5737	&0.4302		&0.0033	&SOPHIE+\\
55852.5084	&0.4398		&0.0034	&SOPHIE+\\
55880.4888	&0.4490		&0.0036	&SOPHIE+\\
55883.5563	&0.4512		&0.0036	&SOPHIE+\\
55955.4716	&0.4693		&0.0037	&SOPHIE+\\
55968.3043	&0.4813		&0.0038	&SOPHIE+\\
55980.3807	&0.4955		&0.0037	&SOPHIE+\\
55983.2592	&0.4855		&0.0031	&SOPHIE+\\
56018.2942	&0.5094		&0.0033	&SOPHIE+\\
56149.6362	&0.5317		&0.0036	&SOPHIE+\\
56234.5222	&0.5402		&0.0034	&SOPHIE+\\
56297.4726	&0.5361		&0.0035	&SOPHIE+\\
56326.2884	&0.5394		&0.0040	&SOPHIE+\\
56361.3075	&0.5455		&0.0022	&SOPHIE+\\
56523.6341	&0.5365		&0.0035	&SOPHIE+\\
56586.6136	&0.5379		&0.0036	&SOPHIE+\\
56631.4207	&0.5456		&0.0035	&SOPHIE+\\
56667.3642	&0.5475		&0.0028	&SOPHIE+\\
56680.3360	&0.5431		&0.0035	&SOPHIE+\\
56733.3712	&0.5342		&0.0028	&SOPHIE+\\
56884.6237	&0.5208		&0.0034	&SOPHIE+\\
56974.6637	&0.4938		&0.0034	&SOPHIE+\\
57000.5342	&0.5060		&0.0036	&SOPHIE+\\
57027.3511	&0.4914		&0.0043	&SOPHIE+\\
57030.4164	&0.4891		&0.0032	&SOPHIE+\\
57030.4229	&0.4927		&0.0017	&SOPHIE+\\
57033.3788	&0.4904		&0.0022	&SOPHIE+\\
57046.3508	&0.4809		&0.0040	&SOPHIE+\\
57051.4553	&0.4756		&0.0044	&SOPHIE+\\
57053.3122	&0.4867		&0.0033	&SOPHIE+\\
57054.3412	&0.4841		&0.0032	&SOPHIE+\\
57055.3789	&0.4803		&0.0044	&SOPHIE+\\
57063.2819	&0.4795		&0.0022	&SOPHIE+\\
57064.3056	&0.4803		&0.0019	&SOPHIE+\\
57065.3232	&0.4806		&0.0021	&SOPHIE+\\
57066.3215	&0.4767		&0.0033	&SOPHIE+\\
57071.3543	&0.4756		&0.0033	&SOPHIE+\\
57072.3032	&0.4748		&0.0033	&SOPHIE+\\
57077.3605	&0.4749		&0.0034	&SOPHIE+\\
57079.4398	&0.4652		&0.0034	&SOPHIE+\\
57081.3098	&0.4716		&0.0033	&SOPHIE+\\
57082.3654	&0.4605		&0.0033	&SOPHIE+\\
57082.3682	&0.4687		&0.0027	&SOPHIE+\\
57119.3022	&0.4458		&0.0029	&SOPHIE+\\
57318.5711	&0.4779		&0.0033	&SOPHIE+\\
57322.5186	&0.4782		&0.0033	&SOPHIE+\\
57341.6912	&0.4961		&0.0020	&SOPHIE+\\
57343.5214	&0.4907		&0.0010	&SOPHIE+\\
57344.5119	&0.4910		&0.0010	&SOPHIE+\\
57345.5194	&0.4907		&0.0011	&SOPHIE+\\
57373.5202	&0.4918		&0.0033	&SOPHIE+\\
57378.3850	&0.4977		&0.0033	&SOPHIE+\\
57383.4111	&0.5038		&0.0016	&SOPHIE+\\
57386.4651	&0.4996		&0.0032	&SOPHIE+\\
57390.4414	&0.5004		&0.0033	&SOPHIE+\\
57394.4581	&0.5211		&0.0050	&SOPHIE+\\
57408.5114	&0.5041		&0.0030	&SOPHIE+\\
57411.3967	&0.5116		&0.0034	&SOPHIE+\\
57411.4415	&0.5011		&0.0034	&SOPHIE+\\
57437.3516	&0.5160		&0.0042	&SOPHIE+\\
57439.2881	&0.5135		&0.0037 &SOPHIE+\\  
\end{longtable}
}
\end{document}